\documentclass[10pt,a4paper,onecolumn]{article}

\pdfoutput=1

\usepackage[margin=1.7cm]{geometry}
\usepackage{amsmath}
\usepackage{amssymb}

\usepackage{graphicx}
\graphicspath{{./figures/}}

\usepackage[T1]{fontenc} %Allows more exotic characters to be displayed properly (eg, Đ) Unclear if needed with LaTeX3
\usepackage{lmodern} %Makes T1 not ugly on a PDF viewer.
\usepackage[super,numbers,comma,sort&compress]{natbib}

\usepackage{titling} % So I can have a second title in the Appendix.
\usepackage{authblk} % To easily add affiliations for authors.

\usepackage{hyperref}
\usepackage[figurename=Fig., labelfont=bf, labelsep=space]{caption} %Mostly here to ensure links to top of figure rather than the caption. Also reformats the caption label.

\usepackage[separate-uncertainty=true,multi-part-units=single]{siunitx}

\newcommand*{\vect}[1]{\mathbf{#1}}
\newcommand*{\bra}[1]{\langle #1 \vert}
\newcommand*{\ket}[1]{\vert #1 \rangle}
\newcommand*{\mean}[1]{\langle #1 \rangle}
\newcommand*{\cre}[2]{\hat{#1}^{\dagger}_{#2}} %Creation operator
\newcommand*{\ann}[2]{\hat{#1}_{#2}} %Annihilation operator
\newcommand*{\numn}[1]{\hat{n}_{#1}} %Number operator, as single operator.
\newcommand*{\numop}[2]{\cre{#1}{#2}\ann{#1}{#2}} %Number operator explicitly expanded.

\renewcommand*{\k}{\vect{k}}

\newcommand*{\vr}{\vect{r}} %Can't use \r, or else \AA (angstrom) doesn't work.

\newcommand*{\eV}{\electronvolt}
\DeclareSIUnit\angstrom{\text{Å}}

\begin{document}

\title{Correlation-induced magnetism in substrate-supported 2D metal-organic frameworks}

\author[1,2]{Bernard Field}
\author[1,2]{Agustin Schiffrin}
\author[2,3,*]{Nikhil V. Medhekar}
\affil[1]{School of Physics and Astronomy, Monash University, Clayton, Victoria 3800, Australia}
\affil[2]{ARC Centre of Excellence in Future Low-Energy Electronics Technologies, Monash University, Clayton, Victoria 3800, Australia}
\affil[3]{Department of Materials Science and Engineering, Monash University, Clayton, Victoria 3800, Australia}
\affil[*]{nikhil.medhekar@monash.edu}
\maketitle

\begin{abstract}
	Two-dimensional (2D) metal-organic frameworks (MOFs) in a kagome lattice can exhibit strong electron-electron interactions, which can lead to tunable quantum phases including many exotic magnetic phases. %26
	While technological developments of 2D MOFs typically take advantage of substrates for growth, support, and electrical contacts, investigations often ignore substrates and their dramatic influence on electronic properties of MOFs. %30
	Here, we show how substrates alter the correlated magnetic phases in kagome MOFs using systematic density functional theory and mean-field Hubbard calculations. % 22
	We demonstrate that MOF-substrate coupling, MOF-substrate charge transfer, strain, and external electric fields are key variables, activating and deactivating magnetic phases in these materials. %24
	While we consider the example of kagome-arranged 9,10-dicyanoanthracene molecules coordinated with copper atoms, our findings should generalise to any kagome lattice. %21
	This work offers useful predictions for tunable interaction-induced magnetism in surface-supported 2D organic materials, opening the door to solid-state electronic and spintronic technologies based on such systems. %27
\end{abstract}
% Latest word count: 150
% Abstract word limit: 150.

\clearpage

\section*{Introduction}

% The first paragraph of the introduction is very important. It usually contains a broad overview of the field of interest and it presents the main challenge which will be addressed by this paper.
Two-dimensional (2D) materials with flat electronic bands, such as the kagome lattice, can host localised electronic wavefunctions, which can lead to strong electron-electron interactions.
These strong interactions lead to diverse electronic phases such as fractional quantum Hall states\cite{tang_high-temperature_2011,sun_nearly_2011}, superconductivity\cite{mazin_theoretical_2014,kiesel_sublattice_2012,kobayashi_superconductivity_2016,aoki_theoretical_2020}, or different magnetic phases, including ferromagnetism\cite{yamada_first-principles_2016}, antiferromagnetism, and spin glasses\cite{pavarini_magnetism_2013,balents_spin_2010}.
Such magnetic phases are critical for emerging technologies such as spintronics and magnetic information storage\cite{klein_probing_2018}.
Kagome lattices which can host such phases have been realised in many different systems.
While inorganic systems such as herbertsmithite and FeSn are well-known examples, a class of materials of growing interest are 2D metal-organic frameworks (MOFs), which combine transition metal ions with planar organic molecules into a self-assembled crystal.
2D MOFs have potential applications in functional electronics, leveraging the intrinsic electrical\cite{sheberla_high_2014}, optoelectronic\cite{sakamoto_photofunctional_2015}, sensing\cite{campbell_cu_3hexaiminotriphenylene_2_2015}, or catalytic\cite{dong_large-area_2015} properties of these systems\cite{zhao_ultrathin_2018,maeda_coordination_2016}, with promise for increased versatility, tunability, and affordability.
For practical solid-state applications, these 2D MOFs are required to interface with substrates and solid interconnects, either for structural support, to enable bottom-up synthesis\cite{barth_molecular_2007,stepanow_modular_2008,goronzy_supramolecular_2018}, or as metallic contacts.
However, the interplay between the various influences of the substrate and the intrinsic properties of the MOF is not understood.

\begin{figure*}
	\centering
	\includegraphics{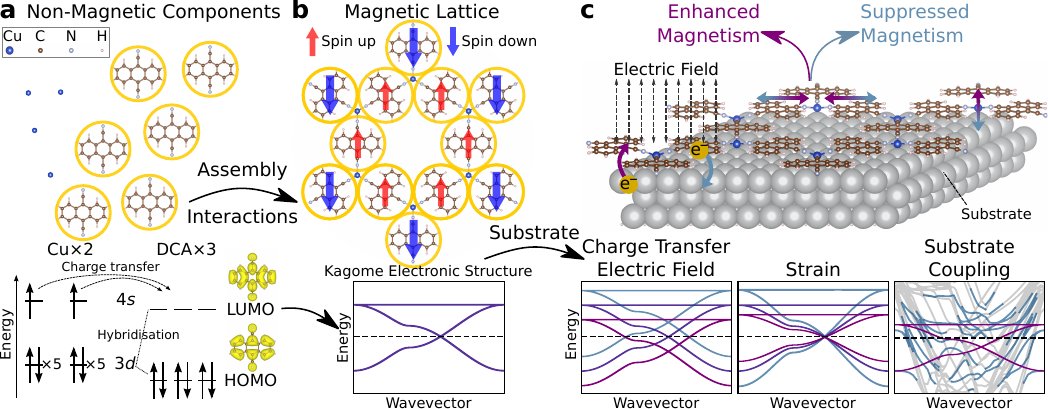}
	\caption{Schematic depiction of interaction-induced magnetism in a 2D MOF and how substrates influence it.
		\textbf{a} Ball-and-stick models of DCA and Cu, the components of the MOF, and their isolated electronic structures. The copper(I) ions and DCA molecules in the MOF are not intrinsically magnetic.
		\textbf{b} Ball-and-stick model of DCA-Cu kagome MOF. After self-assembly, intra-MOF electron-electron Coulomb interactions induce magnetic moments in the MOF. The MOF has a kagome electronic band structure derived from DCA LUMO (non-spin-polarised schematic shown).
		\textbf{c} Ball-and-stick model of DCA-Cu on a substrate and schematic band structures.
		The magnetism in the MOF is influenced by coupling to the substrate, charge transfer either out of or into the MOF (such as by an applied electric field), and strain, which alters the bandwidth.
		The strength of the magnetic moments can be enhanced by using a weakly interacting substrate, favourable electron filling of the MOF by choice of substrate work function or application of electric fields, and by applying tensile strain to the MOF such as by lattice mismatch.
		(HOMO: highest occupied molecular orbital. LUMO: lowest unoccupied molecular orbital)
	}
	\label{fig:summary}
\end{figure*}

% Paragraph purpose: Describe the kagome lattice
The kagome lattice is an archetypal 2D flat-band crystal structure, with a three-site basis arranged in a six-pointed star motif (Fig. \ref{fig:summary}b).
Its electronic band structure has a pair of Dirac bands, similar to graphene.
The frustrated geometry from the triangular three-atom basis allows for destructive interference which localises the electronic wavefunctions and creates an electronic flat band (in addition to the 2 Dirac bands).
This localisation and narrow bandwidth enhances electronic interactions\cite{hubbard_electron_1963}, which can lead to phenomena driven by many-body electron correlations, including magnetism (due to energy cost, from Coulomb interactions, of two electrons occupying the same site exceeding the kinetic energy).
This geometry also allows kagome lattices to host exotic spin behaviour, such as spin liquids, frustrated antiferromagnetism, intrinsic Kondo resonances, and spin density waves\cite{balents_spin_2010,zhang_many-body_2020,hua_highly_2021,kim_emergent_2020}, which can be explored by tuning the interactions\cite{kim_emergent_2020,ferhat_phase_2014}.
Furthermore, as in other Dirac materials, spin-orbit coupling can lead to topological quantum spin Hall phases\cite{wang_quantum_2010}.
The combination of geometry, topology, and interactions makes the kagome lattice a rich playground of electronic quantum phenomena.

% Paragraph purpose: Talk about MOFs.
Of the material systems which kagome lattices have been realised in, an emerging class of materials are 2D MOFs.
In MOFs, organic ligands connect to transition metal atoms via coordination chemistry and form an extended crystal lattice.
By using organic chemistry to alter the ligands and choosing between a large number of transition metal ions, MOFs demonstrate incredible versatility and tunability.
2D MOFs can be electrically conducting, giving them applications in functional electronics\cite{zhao_ultrathin_2018,maeda_coordination_2016}.
MOFs also offer room-temperature synthesis\cite{zhao_ultrathin_2018} which makes them cheaper to produce than inorganic materials.
Kagome MOFs have attracted much attention in the search for topological insulators\cite{wang_computational_2017,liu_flat_2013,su_prediction_2018,baidya_chern_2019}.
However, such searches rarely investigate interactions and the resultant magnetism which can compete with topology\cite{kim_competing_2016}.

% Paragraph purpose: Substrates.
In particular, the competing influences of the substrate to interactions in MOFs are often overlooked.
Substrates support, template, and enable the bottom-up synthesis of 2D MOFs\cite{barth_molecular_2007,stepanow_modular_2008,goronzy_supramolecular_2018}.
These substrates can facilitate new functionalities via effects such as charge doping\cite{yan_synthesis_2021,zhou_stacking_2014}, hybridisation\cite{shao_pseudodoping_2019,dreher_proximity_2021,dendzik_substrate-induced_2017}, strain\cite{roldan_strain_2015}, symmetry breaking\cite{sun_deconstruction_2018}, rearrangement of bonds facilitating magnetism\cite{ramasubramaniam_substrate-induced_2009}, and the Kondo effect\cite{tsukahara_evolution_2011,perera_spatially_2010,girovsky_long-range_2017,mugarza_spin_2011,gao_design_2020,kumar_manifestation_2021}.
Substrates are sometimes deleterious, such as through excessive hybridisation or unfavourable symmetry breaking\cite{sun_deconstruction_2018}, and sometimes beneficial, such as through mediating magnetic interactions\cite{girovsky_long-range_2017,tuerhong_two-dimensional_2018} or favourable charge transfer\cite{crasto_de_lima_quantum_2018}.
Clearly, the influence of substrates on strongly-correlated MOFs or on correlated magnetism is important, but it has not been thoroughly investigated.

% Paragraph purpose: Introduce DCA-Cu as our system of interest and why we choose it as an example.
An ideal example of a correlated kagome MOF is 9,10-dicyanoanthracene-copper (DCA-Cu) --- a 2D MOF consisting of kagome-arranged DCA molecules coordinated with Cu atoms, where each Cu atom binds to 3 DCA molecules in a trigonal planar geometry (Fig. \ref{fig:summary}b).
DCA-Cu has been experimentally synthesised on Ag(111)\cite{kumar_manifestation_2021}, Cu(111)\cite{zhang_probing_2014,hernandez-lopez_searching_2021,pawin_surface_2008}, graphene\cite{yan_synthesis_2021}, and NbSe$_2$\cite{yan_two-dimensional_2021}.
DCA-Cu has clean kagome bands, meaning it is well-described by a simple nearest-neighbour tight-binding model.
DCA-Cu has been predicted to be a topological insulator, with substitution of metal atoms enhancing this effect\cite{zhang_intrinsic_2016}.
DCA-Cu has also been predicted to have strong electron-electron interactions \cite{fuchs_kagome_2020,yan_synthesis_2021}, making it a good system to investigate correlated physics.
A sign of the electronic interactions in DCA-Cu are the magnetic moments it hosts when on Ag(111)\cite{kumar_manifestation_2021}.
Unlike MOFs with intrinsically magnetic components, where the magnetic moments are fixed, the variable magnetism in DCA-Cu arises due to electronic interactions which cause otherwise mobile electrons to become spin-polarised, as described by the Hubbard model\cite{kumar_manifestation_2021}.
However, magnetic moments in DCA-Cu have not been observed on other substrates and it is unclear why some substrates support magnetism in this MOF while others suppress it.

% Paragraph purpose: Present the methods and results of this paper.
Here we show how the coupling between a kagome lattice and a substrate influences electron-electron interactions, leading to the emergence or suppression of magnetic moments and magnetic order.
We investigate the emergent magnetism in DCA-Cu on several substrates (Ag(111), Cu(111), Au(111), Al(111), graphite, and hexagonal boron nitride (hBN)) using a combination of density functional theory (DFT) and a mean-field Hubbard model.
The systematic DFT calculations provide a quantitative description of electronic and magnetic properties of DCA-Cu, including the emergence of magnetically ordered phases.
In particular, we found magnetism in DCA-Cu on Ag(111), DCA-Cu on hBN on Cu(111), DCA-Cu on graphite with tensile strain, and DCA-Cu on Cu(111) with an electric field.
Importantly, we found ways of controlling such magnetism (i.e., switching on and off) via external stimuli, namely applied electric field and strain.
Meanwhile, we present the Hubbard Hamiltonian as a simple model which captures the key features of the DFT calculations and allows for interpreting the results.
The Hubbard model distils these complicated systems into a small number of key ingredients, including MOF-substrate coupling, while being generic to any kagome lattice on a substrate.
Both DFT and Hubbard approaches consistently show that certain substrate parameters, namely the density of states, MOF-substrate charge transfer, MOF-substrate electronic hybridisation, and substrate-induced MOF tensile strain, can dramatically affect the MOF magnetic properties (Fig. \ref{fig:summary}c).
Our work quantitatively isolates the effect of each of these phenomena on the system's magnetic properties.
Sensitivity to charge transfer allows switching the magnetic phases by an external electric field, which we explicitly demonstrate in DCA-Cu, suggesting possible applications in energy-efficient electric-field--controlled spintronics\cite{ramesh_electric_2021,matsukura_control_2015}.
% Paragraph purpose: Speculative significance of this work.
Understanding how different parameters affect the magnetism in this MOF allows us to tune the electron-electron interactions and magnetic phases, a feature not offered by intrinsically magnetic systems. By developing a framework for analysing the effects of a substrate on an adsorbed 2D system, we present useful tools for the rational design of novel electronic devices whose functionality relies on correlations-induced quantum phenomena.

\begin{figure*}
	\centering
	\includegraphics{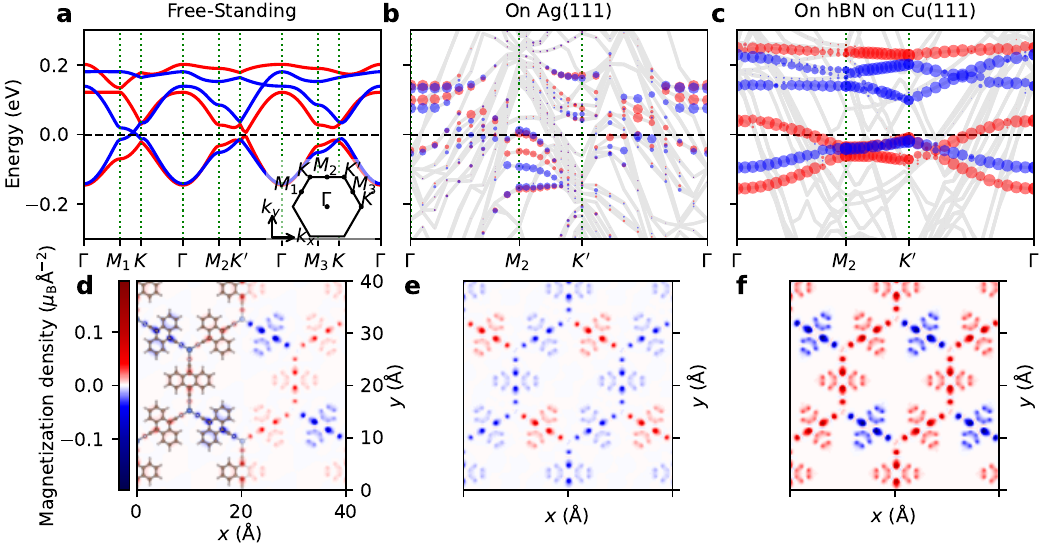}
	\caption{DFT+$U$ band structures and magnetization densities of DCA-Cu (\textbf{a,d}) without a substrate, (\textbf{b,e}) on Ag(111), and (\textbf{c,f}) on hBN on Cu(111).
		Strong electron-electron Coulomb interactions separate spin up and down bands (red and blue respectively) and lead to local magnetic moments on DCA molecules.
		On Ag(111) (\textbf{b}), substrate coupling distorts the bands, while on hBN (\textbf{c}) coupling is negligible so bands retain free-standing character (\textbf{a}). Circle radius in band structures is proportional to projection of bands onto the atoms of DCA-Cu. Inset to \textbf{a} is the Brillouin zone.
		Overlaid in \textbf{d} is the atomic structure of free-standing DCA-Cu.}
	\label{fig:bands}
	\label{fig:spin-density}
\end{figure*}

\begin{figure}
	\centering
	\includegraphics{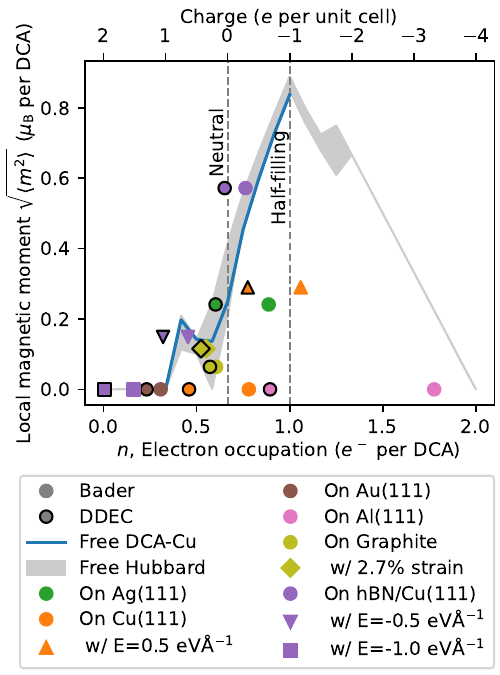}
	\caption{Charge transfer and local magnetic moments of DCA-Cu MOF (free-standing and on substrates) and under different conditions, calculated using DFT+$U$.
		These local magnetic moments lead to magnetically ordered phases (Figs. \ref{fig:spin-density} and \ref{fig:hubbard-un}).
		Two different charge partitioning methods set bounds for the MOF-substrate charge transfer: Bader\cite{tang_grid-based_2009} and DDEC\cite{manz_introducing_2016,limas_introducing_2016}.
		A third method, integrating the projected density of states, gives values between the Bader and DDEC results (Supplementary Fig. \ref*{fig:chg-mag-u0}).
		E refers to the electric field applied normal to the substrate.
		The grey shaded region marks the results calculated within the mean-field Hubbard model with a single unit cell, without a substrate, and with interaction parameter $U_H$ between $6t$ and $7t$, presented for comparison with the DFT-calculated free-standing results.}
	\label{fig:chg-mag-summary}
\end{figure}

\section*{Results}

\subsection*{Electronic and magnetic structure of DCA-Cu}

% Paragraph purpose: DFT+U and intro
To understand the effect of real substrates on real materials, we use DCA-Cu as an illustrative example and perform systematic DFT+$U$ calculations for free-standing DCA-Cu as well as for different substrates (see Methods).
The use of Hubbard $U$ corrections treats the correlations in $d$ electrons and are essential for an accurate description of magnetic moments in DCA-Cu on Ag(111), without which there are no magnetic moments when experiments show there should be\cite{kumar_manifestation_2021}.
We found that the effect of $U$ on the Cu atoms is to slightly reduce the hopping between DCA molecules, which is mediated by the Cu atoms (see Supplementary Table \ref*{tab:dft-mfh-u-fitting}).
We report results for $U=\SI{3}{\eV}$; besides the Ag(111) results, varying $U$ within a reasonable range does not cause any qualitative changes (see Supplementary Fig. \ref*{fig:chg-mag-u0} and Ref. \cite{kumar_manifestation_2021}).
From these calculations, we determine which substrates enhance or suppress local magnetic moments in DCA-Cu and identify the contribution of charge transfer.
We also address how an externally applied electric field affects these local magnetic moments.

% Paragraph purpose: Describe electronic structure of freestanding DCA-Cu
We first investigate the electronic structure of free-standing DCA-Cu without a substrate.
For calculations without spin polarisation, the band structure of free-standing DCA-Cu matches an ideal nearest-neighbour tight-binding kagome band structure (Fig. \ref{fig:summary}b and Supplementary Fig. \ref*{fig:dft-tb-comparison}a).
Electronically, each Cu atom donates one electron to the DCA lowest unoccupied molecular orbitals (Fig. \ref{fig:summary}a), which form the kagome bands at the Fermi energy.
This integer charge transfer pins the Fermi level at the Dirac point and leaves the Cu$^+$ ions in the nominally non-magnetic [Ar]$3d^{10}$ configuration.
When spin is included, electron-electron interactions lead to spin-splitting of bands and magnetization of DCA orbitals (Fig. \ref{fig:bands}a,d).
The system prefers a frustrated antiferromagnetic configuration by about \SI{1}{\milli\eV (Supplementary Table \ref*{tab:spin-summary})}.
This magnetization cannot be explained by simple counting of unpaired spins as the magnetic moments are a fraction of a Bohr magneton, instead requiring electronic interactions as in the Hubbard model to partially hinder electron pairing\cite{kumar_manifestation_2021}.
As we show next, these interactions are altered drastically when the 2D MOF is in contact with a substrate.

\subsection*{Emergent magnetism in DCA-Cu on substrates}

% Paragraph purpose: Introduce and justify our selection of substrates.
To elucidate the substrate's influence on the electronic and magnetic structure of DCA-Cu, we sample a broad selection of different substrates.
We choose substrates with a range of work functions and lattice constants and a mixture of metals and gapped materials (Supplementary Table \ref*{tab:substrates}).
For metals, we consider Ag(111), Cu(111), Au(111), and Al(111), which are structurally similar but have differing work functions.
We consider graphite, which is a semimetal so has a much lower density of states and can be regarded as multi-layer graphene in this context.
And we consider hexagonal boron nitride (hBN), a wide band gap insulator commonly used in 2D heterostructures with potential for electronic devices, on Cu(111), a common surface for growing hBN\cite{wang_recent_2017}.
% hBN acronym is already defined in the introduction.
While not a comprehensive survey of all possible substrates, this selection represents a sufficient sample size for identifying the essential trends.

% Paragraph purpose: Ag(111) results
We find that of the six substrates considered, only two retain local magnetic moments in DCA-Cu, as shown in Fig. \ref{fig:spin-density}.
Our calculations indicate that these systems retain some degree of frustrated antiferromagnetic order between such local magnetic moments (albeit with some net magnetization resembling ferrimagnetism; see Supplementary Table \ref*{tab:spin-summary}).
One of them is DCA-Cu on Ag(111), in which previous experiments\cite{kumar_manifestation_2021} showed direct evidence of local magnetic moments --- consistent with our calculations --- via the observation of the Kondo effect (i,e,, where such local magnetic moments are screened by the Ag(111) conduction electrons).
The band structure of DCA-Cu on Ag(111) shows substantial distortion of the kagome bands, indicating non-negligible MOF-substrate coupling (Fig. \ref{fig:bands}b).
In spite of this, magnetization of DCA orbitals persists on Ag(111) (Fig. \ref{fig:spin-density}e) with stability comparable to free DCA-Cu at \SI{1.5}{\milli\eV} (Supplementary Table \ref*{tab:spin-summary}).
It is important to note, however, that such magnetic order might not be realised experimentally on a metal such as Ag(111) due to the aforementioned Kondo screening, whose many-body physics are not captured by our DFT approach.

% Paragraph purpose: hBN results
The other substrate to retain magnetic moments is hBN on Cu(111).
The kagome band structure does not exhibit any avoided crossings, behaving similarly to free-standing DCA-Cu (Fig. \ref{fig:bands}c, Supplementary Fig. \ref*{fig:dft-tb-comparison}b).
Compared to free-standing DCA-Cu, the shifted Fermi level indicates that DCA-Cu receives electrons from the substrate, and also has local magnetic moments with greater magnitude (Fig. \ref{fig:spin-density}f).
The magnetic moments are also more stable, with the magnetic configuration \SI{23.9}{\milli\eV} lower than the nonmagnetic configuration (Supplementary Table \ref*{tab:spin-summary}).
Note that these configurations are constrained to using a single kagome unit cell.
Our Hubbard model calculations in a supercell (next section) indicate that more stable configurations have a larger magnetic unit cell (if they are ordered at all).
This suggests that the local magnetic moments may be even more stable than single-cell DFT calculations indicate.
That DCA-Cu/hBN/Cu(111) has a higher local magnetic moment than DCA-Cu/Ag(111) suggests that reduced MOF-substrate hybridisation enhances the local magnetic moments.
However, the difference between DCA-Cu/hBN/Cu(111) and free-standing DCA-Cu needs to be explained with other variables.

% Paragraph purpose: results of the other substrates
The other substrates which were considered --- graphite, Cu(111), Au(111), and Al(111) --- all suppress the local magnetic moments in DCA-Cu (Supplementary Fig. \ref*{fig:misc-bands}).
This observation is consistent with experiments of DCA-Cu on Cu(111), which also did not find magnetic moments\cite{kumar_manifestation_2021,zhang_probing_2014}.
The avoided crossings in the band structures indicate that Cu(111), Au(111), and Al(111) are strongly interacting (just like Ag(111)) while graphite is weakly interacting.
The bands also clearly indicate that Au(111) fully depopulates the DCA-Cu kagome bands, Al(111) fully populates the DCA-Cu kagome bands, and graphite slightly reduces the electron occupation of DCA-Cu.

This raises the important question: why do some substrates allow magnetism in the MOF while others do not?
Figure \ref{fig:chg-mag-summary} shows the influence of charge transfer between the substrate and MOF on the local magnetic moments on DCA.
Free-standing DCA-Cu calculated by both DFT and the mean-field Hubbard model with varying charge is also included as a control.
Determining an unambiguous charge transfer can be challenging due to electron density partitioning effects.
Therefore we use multiple methods to estimate plausible bounds for the charge transfer (see Methods and Supplementary Notes).
Free-standing DCA-Cu shows that local magnetic moments are maximised when the orbitals are half-filled (one electron per orbital) and weaken rapidly away from half-filling.
This trend is corroborated on substrates.
Substrates with electron fillings closer to half-filling (Ag(111), hBN/Cu(111)) show magnetic moments, while substrates which deplete electrons (Cu(111), Au(111)) or completely fill the bands (Al(111), see also Supplementary Fig. \ref*{fig:misc-bands}c) do not show any magnetic moments.
Graphite, while having much weaker MOF-substrate coupling than bare metals, also has very weak magnetic moments due to a low electron occupation.
Overall, it is clear that electron occupation of the MOF is a significant factor in determining whether or not magnetic moments occur in the MOF.
But it is also clear that the magnetic moments deviate from the free-standing results when a substrate is considered.
This indicates that variables such as MOF-substrate hybridisation and/or substrate-induced strain also influence the magnetic moments.
To analyse these variables we turn to a simplified model.

\begin{figure}
	\centering
	\includegraphics{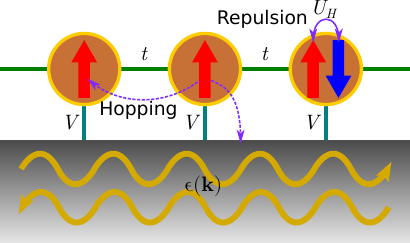}
	\caption{Schematic representation of our physical model of emergent magnetism in substrate-supported 2D MOFs.
	The 2D lattice is treated within the Hubbard model with nearest-neighbour hopping $t$ (tunable by strain) and Coulomb repulsion $U_H$.
	The substrate is modelled as having plane wave states with dispersion $\epsilon(\k)$ (derived from \textit{ab initio} calculations).
	Coupling between the substrate and the lattice is assumed to be point-like hopping in real-space, with strength $V$.
	The lattice sites have an on-site energy, which alters the electron occupation of the lattice based on its alignment with the substrate chemical potential.}
	\label{fig:model}
\end{figure}

\begin{figure*}
	\includegraphics{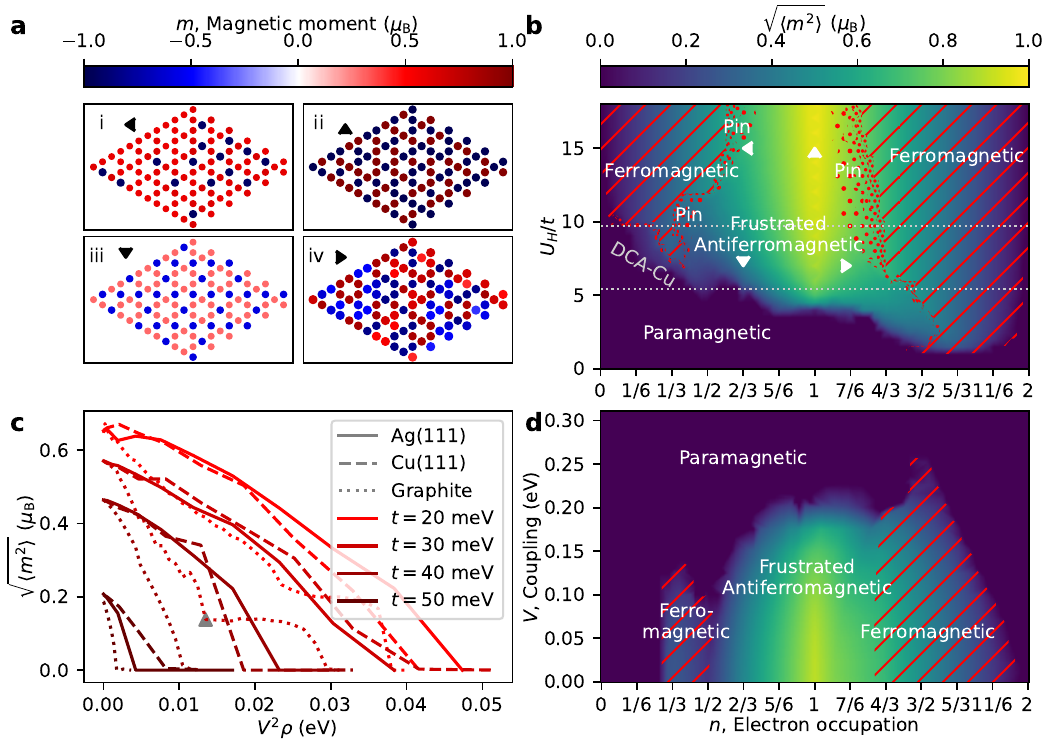}
	\caption{Emergent magnetism in the kagome lattice from our mean-field Hubbard model.
	\textbf{a} Real-space magnetization densities in a free-standing kagome lattice.
	We show a pinned metal droplet phase\cite{kim_emergent_2020} (i), a spin-density-wave phase (iii), and two disordered phases (ii, iv).
	\textbf{b} Magnetic phase diagram of the free-standing kagome lattice in a supercell.
	The red dotted region indicates the transition from antiferromagnetic to ferromagnetic phase, with a partial net polarisation (Supplementary Figure \ref*{fig:hubbard-un-mag}).
	`Pin' refers to a pinned metal droplet phase.
	Triangular markers indicate phase space locations corresponding to \textbf{a}.
	Dotted horizontal lines indicate the range of $U_H/t$ values that reproduce the DFT results for DCA-Cu (Supplementary Table \ref*{tab:dft-mfh-u-fitting} and Supplementary Figure \ref*{fig:dft-mfh-bands}).
	\label{fig:hubbard-un}
	\textbf{c} Effect of effective hybridisation $V^2\rho$ and MOF hopping parameter $t$ (proportional to lattice bandwidth) on the strength of magnetic moments for different substrates.
	The substrate density of states $\rho$ is taken as the mean DOS of the substrate dispersions over \SI{+-0.2}{\eV} around the Fermi level which is 0.76, 0.82, and \SI{0.035}{\per\eV} for Ag(111), Cu(111), and graphite, respectively.
	The kink in the \SI{30}{\milli\eV} graphite curve is due to an antiferromagnetic to ferromagnetic phase transition (see Supplementary Note \ref*{sec:afm-fm}).
	Data obtained using $U_H=\SI{300}{\milli\eV}$ and an electron occupation of approximately 2/3.\label{fig:hubbard-substrate-varyt}\label{fig:hubbard-substrate-agvscu}\label{fig:hubbard-dos}
	\textbf{d} Effect of electron occupation and coupling to a substrate on the strength of magnetic moments.
	Magnetic phases can withstand higher coupling with the substrate at higher electron occupations than lower electron occupations.
	Data obtained using a single kagome unit cell, $U_H=\SI{300}{\milli\eV}$, $t=\SI{40}{\milli\eV}$, and the Ag(111) dispersion.\label{fig:hubbard-substrate-coupling-vs-n}}
	\label{fig:mfh}
\end{figure*}

\subsection*{Mean-field Hubbard model for emergent magnetism in a kagome lattice on a metal}

% Paragraph purpose: Justify MFH, give an overview of the method.
To explore and rationalise the essential ingredients which lead to the physics observed via the DFT calculations, we develop a simplified mean-field model (Fig. \ref{fig:model}).
We model the kagome MOF as a tight-binding lattice with nearest-neighbour hopping $t$ and on-site Coulomb interactions $U_H$, as in the Hubbard model\cite{hubbard_electron_1963}.
In this model, electrons hop between adjacent sites due to $t$ and are repelled if they share a site with another electron due to $U_H$, which will be of the opposite spin due to the Pauli exclusion principle.
Note that the $U_H$ used here is different from $U$ in the DFT+$U$ calculations; the former describes all electronic interactions including the DCA molecular orbitals, while the latter is a correction to electronic correlations in Cu $d$ orbitals.
In our model, electrons in the substrate propagate as plane waves with periodic energy dispersion $\epsilon(\k)$ which we derive from DFT calculations for metal slabs.
Electrons can hop between the point-like MOF sites and the substrate with coupling amplitude $V$, as in the Anderson impurity model\cite{anderson_localized_1961}.
We can use supercells to analyse order at larger distances than in DFT.
Unlike DFT, we can vary all the parameters independently and continuously.
Considering the connection between this model and a physical system, $t$ is determined by bonding between molecules within the 2D kagome system, $V$ is determined by bonding between the kagome MOF molecules and the substrate, and $U_H$ is determined primarily by the localisation of the 2D kagome MOF molecular orbitals and their dielectric environment.
This simplified model allows for capturing and exploring different magnetically ordered phases, interpreting our \textit{ab initio} DFT results, and identifying general trends in parameter space.

% Paragraph purpose: Explain the free-standing results and how U, t, and n influence magnetism.
First we demonstrate the emergent magnetism in a free-standing kagome lattice using our Hubbard model.
As shown in Figure \ref{fig:hubbard-un}a-b, when $U_H$ is small compared to the bandwidth ($6t$) the system is paramagnetic, while the local magnetic moments have large magnitude when $U_H$ is large or $t$ is small.
The local magnetic moments are maximal near half-filling ($n=1$) with whole spins sitting on each site in an Ising-like configuration.
The frustrated antiferromagnetic phase consists of intricate spin textures, as indicated in Fig. \ref{fig:hubbard-un}a, which arise due to geometric frustration from the kagome lattice and energetically favourable antiferromagnetic interactions.
Some of these spin textures (iii) are periodic spin-density-wave phases\cite{brown_charge_1994,kim_competing_2016}, while others (ii,iv) are disordered, potentially resembling spin liquids\cite{balents_spin_2010} or spin ices.
Precise classification of disordered magnetic phases in the kagome lattice is beyond the scope of this study; we focus instead on the strength of individual magnetic moments.
Away from half-filling the system transitions towards a ferromagnetic phase, passing an intermediate, partially polarised phase with pinned minority spins\cite{kim_emergent_2020} (Fig. \ref{fig:hubbard-un}a(i)).
See Supplementary Figure \ref*{fig:hubbard-un-mag} for corresponding net magnetization.
Whether these phases have some itinerant nature or not is beyond the scope of this work and requires deeper investigation.
Fundamentally, larger interactions $U_H$, smaller hopping $t$, or electron occupations $n$ closer to one per site enhance the magnetic moments.
Stronger electron-electron interactions within the kagome lattice lead to less pairing of electron spins, and electron occupations closer to one per site allow more unpaired electrons.

% Paragraph purpose: Explain how V affects magnetism.
With the substrate-free results as a baseline, we next investigate magnetism in a kagome lattice coupled to metallic substrates, specifically Ag(111), Cu(111), and graphite (Fig. \ref{fig:hubbard-substrate-agvscu}c).
Note that, in these calculations, the use of these substrates is purely for their electronic dispersion (calculated by DFT for a thin slab), while $V$ is (for now) a free parameter.
Figure \ref{fig:hubbard-substrate-agvscu}c shows the local magnetic moments of kagome lattices with different hopping $t$ coupled to different substrates with respect to the effective hybridisation between the substrate and the lattice.
This effective hybridisation can be expressed, to first order, as $V^2\rho$, where $\rho$ is the substrate density of states (DOS) near the Fermi level, similar to the Anderson impurity model\cite{anderson_localized_1961} (also see Supplementary Fig. \ref*{fig:hubbard-triangle-dos}).
When this effective hybridisation is increased, either with larger MOF-substrate coupling $V$ or higher DOS, the local magnetic moments decrease until the system becomes paramagnetic above some critical hybridisation.
Small deviations between substrates come from finer details of the band structure, such as non-uniform DOS over the kagome bandwidth.
Nevertheless, this scaling brings substrates with very different DOS to similar values, indicating that it captures the main contributions ---
greater coupling with the substrate or higher substrate DOS reduces the local magnetic moments in the kagome lattice.
% Paragraph purpose: Explain how t affects critical coupling.
It is also evident from Fig. \ref{fig:mfh}c that increasing intra-kagome hopping $t$ lowers the critical hybridisation at which the system becomes paramagnetic.
As the hopping between sites is proportional to the overlap between the localised orbitals and is inversely proportional to the distance between molecules, this observation points to a tantalising prospect for tuning the magnetic phases with externally induced strain, as we show in the next section.

% Paragraph purpose: Explain how n affects critical coupling.
We next investigate the influence of electron occupation on the magnetic moments, as shown in Fig. \ref{fig:hubbard-substrate-coupling-vs-n}d.
Reducing the electron occupation significantly below half-filling ($n=1$) substantially lowers the critical coupling (the transition from magnetic to paramagnetic).
The critical coupling does not decrease when $n$ is increased above half-filling until the bands are almost entirely filled, a difference which may come from particle-hole asymmetry.
Figure \ref{fig:mfh}d does not contain a `pin' phase like Fig. \ref{fig:mfh}b because it was obtained with a single kagome unit cell rather than a supercell for computational ease.
See Supplementary Fig. \ref*{fig:hubbard-un-single} for the phase diagram of a free-standing single cell for a more direct comparison.
Overall, electron occupation is an important variable for controlling local magnetic moments, which can be tuned by choice of substrate work function or application of electric fields (i.e. electrostatic gating), as we illustrate in the next section.

% Parameter estimations.
We estimate values for parameters in our mean-field Hubbard model from the DFT calculations.
The hopping parameter $t$ can be obtained from the bandwidth of non-spin-polarised free-standing DCA-Cu, giving $t=\SI{49.2}{\milli\eV}$, although this varies on substrates with different lattice constants as described in the next section.
Interactions can be estimated from fitting the magnetic moments between DFT and mean-field Hubbard results.
We obtained $U_H\simeq 0.29-\SI{0.35}{\eV}$ by fitting the magnetic moments between DFT and mean-field Hubbard results for free-standing DCA-Cu (Supplementary Table \ref*{tab:dft-mfh-u-fitting}).
Electron occupation is obtained directly from DFT (Fig. \ref{fig:chg-mag-summary}).
MOF-substrate coupling can be estimated by fitting tight-binding band structures ($U_H=0$) with non-spin-polarised DFT band structures (Supplementary Fig. \ref*{fig:tb-fits}).
While this fitting is very coarse due to simplified modelling of the substrate in our toy model, on Ag(111) and Cu(111) we obtained $V\simeq \SI{0.2+-0.05}{\eV}$.
It should be emphasised that when parameters are judiciously chosen near these values, the mean-field Hubbard model can reproduce the magnetism (Supplementary Fig. \ref*{fig:mfh-chosen-parameters}), magnetization densities, and band structures (Supplementary Fig. \ref*{fig:dft-mfh-bands}) obtained using DFT calculations.
A good first estimate of the parameters can be obtained from simpler DFT calculations.
Considering the full system only causes a modest (although noticeable) renormalization of the parameters.
We demonstrate this by comparing the spin-resolved band structures of both free-standing DCA-Cu and DCA-Cu on Ag(111) obtained using DFT and the Hubbard model in Supplementary Figure \ref*{fig:dft-mfh-bands}.
Obtaining accurate, complicated observables (e.g. spin-resolved band structure) using a small number of parameters determined by fitting relatively simple quantities (here, bandwidth or magnetization density) demonstrates that our mean-field Hubbard model captures the essential electronic and magnetic properties of the system.
These comparisons show that our simple mean-field Hubbard model fully captures the physics seen in DFT calculations.
The compellingly validates and establishes the relevance of the Hubbard model to quantitatively describe the electronic and magnetic properties of the systems considered.

\begin{figure*}
	\includegraphics{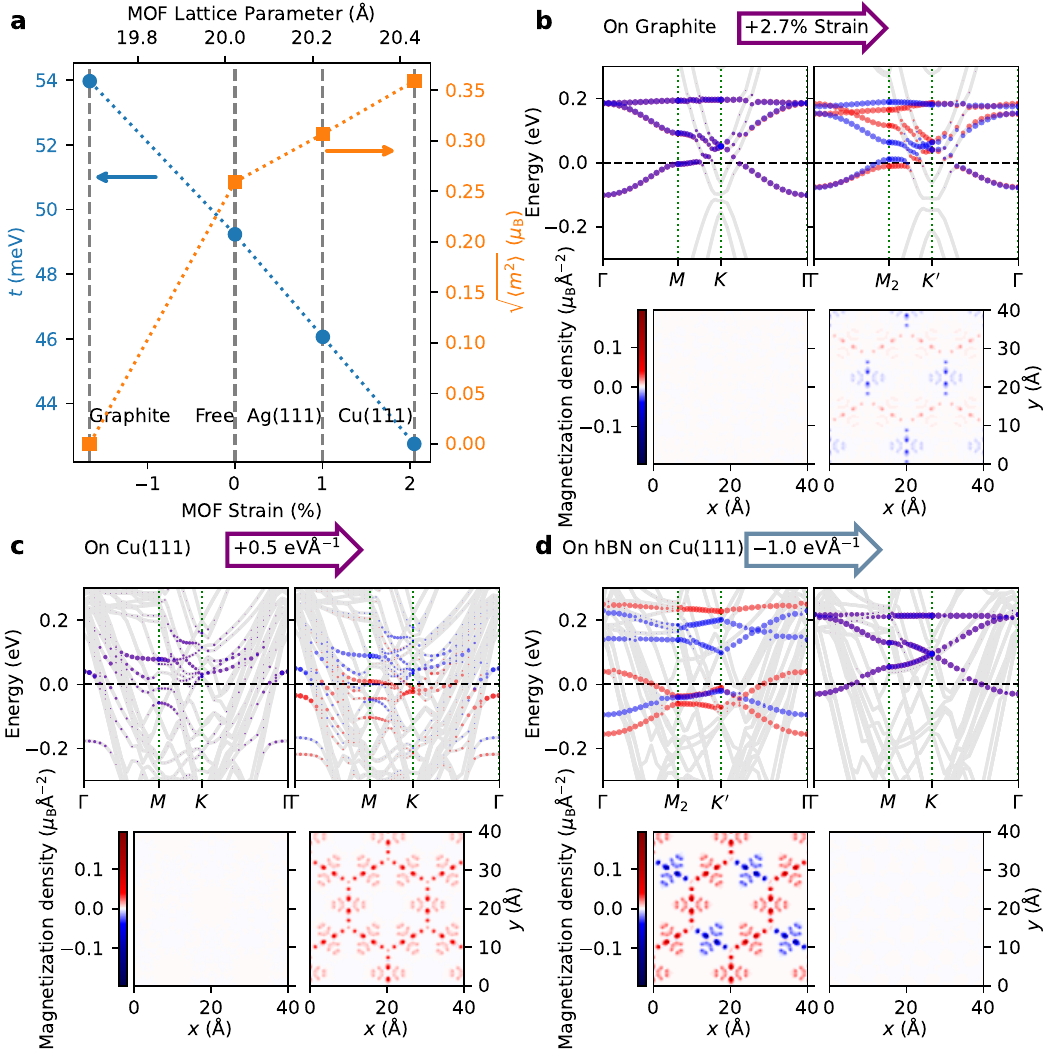}
	\caption{Tailoring magnetism in DCA-Cu on substrates by strain and electric fields.
		\textbf{a} Variation of hopping constants and local magnetic moments of free-standing DCA-Cu due to strain.
		Vertical lines mark lattice constants of different substrate supercells for reference, although these DFT calculations do not include substrates.
		\textbf{b} Band structure and magnetization density of DCA-Cu on graphite before and after applying strain.
		\textbf{c} Band structure and magnetization density of DCA-Cu on Cu(111) before and after applying an electric field normal to the plane pointed towards the substrate, increasing electron filling.
		\textbf{d} Band structure and magnetization density of DCA-Cu on hBN on Cu(111) before and after applying an electric field in the opposite direction, decreasing electron filling.
		Legends in \textbf{b-d} have the same meaning as in Fig. \ref{fig:bands}.
		Purple circles are spin degenerate.}
	\label{fig:strain}
	\label{fig:efield}
\end{figure*}

\subsection*{Tunable control over emergent magnetism: strain and electric fields}

% Paragraph purpose: Introduce strain as something to care about.
We have shown that the strength of the local magnetic moments in 2D MOFs depends strongly on the molecule-molecule hopping $t$ and the electron occupation in MOFs.
Here we demonstrate with DFT how we can exploit this knowledge for tunable external control over the magnetism in a substrate-supported 2D MOF.
First, the mean-field Hubbard model showed that decreasing molecule-molecule hopping $t$ increases the local magnetic moments.
Hopping is proportional to orbital overlap, which is reduced by increasing the intermolecular distance and lattice constant, that is, by applying tensile strain.
We consider in-plane strain here, which affects in-plane hopping $t$.
Applying strain out-of-plane might be used to tune $V$ (which would depend on molecule-substrate overlap), although is beyond the scope of this work.
In-plane strain in a 2D MOF can be introduced by either lattice mismatch with the substrate or by mechanically straining the MOF-substrate system.

% Paragraph purpose: Show effect of strain on free-standing DCA-Cu.
Using DFT, we consider how in-plane strain affects $t$ and local magnetic moments for free-standing DCA-Cu in Fig. \ref{fig:strain}a.
As expected, tensile strain reduces the bandwidth and hopping parameter $t$, increasing the local magnetic moments.
For example, 2\% tensile strain reduces $t$ in free-standing DCA-Cu from 49 to \SI{43}{\milli\eV} and increases the average magnetic moment from 0.26 to 0.36$\mu_B$.
Conversely, compressive strain increases $t$ and causes the system to become paramagnetic.

% Paragraph purpose: Show effect of strain on DCA-Cu/Graphite.
We illustrate how externally applied in-plane strain can be used to activate magnetism in a substrate-supported MOF where magnetism would not otherwise be present.
Consider a graphite substrate, which in its equilibrium state exerts compressive strain on DCA-Cu due to lattice mismatch.
Due to this substrate-induced compressive strain, DCA-Cu on graphite has considerably weaker local magnetic moments than in its free-standing state (Fig. \ref{fig:chg-mag-summary}).
If we apply tensile strain, e.g. 2.7\%, to the DCA-Cu/graphite system the local magnetic moments double (Fig. \ref{fig:strain}b).
This demonstrates that strain from either lattice mismatch or external means is an important factor in tailoring the local magnetic moments in 2D MOFs.

%\subsection{Electric Fields}

The results presented earlier showed that the strength of the local magnetic moments depended strongly on the electron occupation of DCA.
This electron occupation can be altered by charge transfer to or from the substrate, either by choice of a substrate with a suitable work function or by applying an external electric field.
By considering two different substrates, namely Cu(111) and hBN on Cu(111), we illustrate how electric fields can alter the magnetic moments.

Earlier, we showed that DCA-Cu on Cu(111) does not have any magnetism due to its low electron occupation.
However, when we apply an electric field of \SI{0.5}{\eV\per\angstrom} towards the substrate, the direction of charge transfer is reversed and DCA-Cu undergoes a magnetic phase transition as shown in Fig. \ref{fig:efield}c, with local magnetic moments similar to DCA-Cu on Ag(111) (Fig. \ref{fig:chg-mag-summary}).
While DCA-Cu favours a ferromagnetic configuration over a frustrated antiferromagnetic configuration in this case (by \SI{0.4}{\milli\eV}; and by \SI{1.4}{\milli\eV} over paramagnetism), this may be due to substrate-mediated coupling of local magnetic moments (Supplementary Note \ref*{sec:afm-fm}).
Conversely, we can also use electric fields to drive a transition towards paramagnetism.
By applying a sufficiently strong electric field to DCA-Cu/hBN/Cu(111) in the opposite direction, the MOF is depleted of charge (Fig. \ref{fig:chg-mag-summary}) and the local magnetic moments vanish.
These results are remarkable because electric fields can be easily applied in a transistor device and do not rely on altering intrinsic material parameters.
The electric field allows for an external control knob that tunes the local magnetization and magnetic moment ordering, opening the doors for control of magnetic phases in 2D kagome materials by electrostatic gating.

% Add a line on screening effects of substrate.

\section*{Discussion}

% Paragraph purpose: Generalisability beyond DCA-Cu.
Our results, using DFT and a mean-field Hubbard model, show that substrate-induced electron occupation, strain, and hybridisation control the emergent magnetism in strongly correlated 2D MOFs.
We found magnetism in several systems: free-standing DCA-Cu, DCA-Cu on Ag(111), DCA-Cu on hBN on Cu(111), DCA-Cu on graphite with tensile strain, and DCA-Cu on Cu(111) with an electric field.
While we have used DCA-Cu as an illustrative example in our DFT calculations, the consistency between DFT and our more general Hubbard model indicates that our results can be generalised to other kagome MOFs.
By choice of the MOF metals and ligands, of the substrate, and of external controls, it should be possible to explore more of the phase space than that considered in this work.
% Paragraph purpose: Consideration of different MOFs.
For example, changing the components of the MOF will change hopping parameter $t$ (e.g. with longer ligands, or tuning metal-ligand coordination strength via choice of functional groups and metal atoms) and on-site interaction $U_H$ (e.g. with more localised orbitals).
The intrinsic electron occupation of the MOF can be altered by electrochemical doping\cite{kambe_-conjugated_2013,kambe_redox_2014}.
These parameters can be calculated with a free-standing DFT calculation, which is far cheaper than a calculation involving the substrate.
The effect of a substrate can then be included in a simple effective model, at least at a qualitative level.
The renormalization of free-standing parameters when a substrate is introduced makes quantitative predictions more challenging, especially in highly sensitive regions of a parameter space.
However, such models might yield quantitative results when considering a supercell (as opposed to a primitive unit cell), or when estimating parameters within a higher level of theory.
This would involve fitting the parameters to quantities or observables in a small-scale \textit{ab initio} calculation, then using those parameters in a larger scale calculation with the effective model.
The effective model could be our mean-field Hubbard model, an extension to the Hubbard model (e.g. spin-orbit coupling, longer range interactions), or the Hubbard model with a more advanced treatment of correlations (e.g. dynamical mean-field theory).
Such modelling allows for easily exploring a wide range of chemical motifs.

% Paragraph purpose: Consideration of which substrate to choose.
As we have shown, the substrate influences the emergent magnetism in 2D MOFs by strain, charge transfer, and hybridisation.
Consequently, a substrate can be chosen to provide the desired strain (by lattice mismatch) and charge transfer (estimated by work function differences).
Some layered substrates, such as hBN on Cu(111), exhibit long-ranged modulation of the work function due to a Moire pattern\cite{joshi_boron_2012}, which could result in spatial modulation of the electron occupation and magnetic phase.
For substrates with strong hybridisation with MOFs, it may be possible to reduce the coupling strength by adding decoupling functional groups to the MOF ligands (e.g. Ref. \cite{vives_synthesis_2009}).
By understanding such trends, we can rapidly narrow the search space of candidate substrates, aiding rational design.

% Paragraph purpose: Consideration of external control knobs.
We have also shown that we can adjust the emergent magnetism by external controls such as electric fields and mechanical strain.
Electrostatic gating could be used to change the charge transfer, potentially allowing for switchable magnetic phases.
Our results indicate that an electron occupation change of approximately 1 electron per unit cell, or \SI{2.5e13}{\per\cm\squared}, is needed to change the magnetic phases.
If we consider a \SI{10}{\nano\meter} thick parallel-plate capacitor with hBN as a dielectric with a dielectric constant of 3.76\cite{laturia_dielectric_2018}, we get a switching voltage of \SI{12}{\volt} and, for a 10 by \SI{10}{\nano\meter} device, a switching energy of \SI{24}{\atto\joule}.
These voltages are experimentally feasible and the energies are very low\cite{ramesh_electric_2021,matsukura_control_2015}.
If higher electric fields or capacitance is required, electrolytic gates could be employed\cite{kim_electrolyte-gated_2013}.
Applied strain, such as from flexible, piezoelectric, or thermally expanding substrates\cite{roldan_strain_2015}, can also activate or suppress magnetic phases.
These controls could be used to induce magnetism in otherwise non-magnetic MOFs or to dynamically switch magnetism on and off.

% Paragraph purpose: Discuss approximations and limitations.
It is worth pointing out that our calculations capture electron correlations at the mean-field level.
The mean-field approximation allows for exploring magnetic phases in the kagome lattice\cite{ferhat_phase_2014,kim_emergent_2020}.
However, effects such as the Kondo effect\cite{kumar_manifestation_2021} or Mott transition are not captured in our calculations.
Higher levels of theory, such as dynamical mean-field theory, could capture these effects and reveal further interesting physics\cite{georges_strongly_2004}.
An accurate description of many-body effects such as the Kondo effect would be important for quantitative comparisons with experiments, including magnetic order or suppression of the latter, beyond the prediction and corroboration of local magnetic moments.
Our DFT and mean-field results suggest some degree of magnetic order on substrates that include metals (e.g., Ag(111)), where Kondo-screening of local magnetic moments can actually prevent such magnetic order\cite{coleman_heavy_2015}.
%Ask editor: What is proper way to cite this book chapter coleman_heavy_2015
Regardless, our approach is useful for predicting the presence of local magnetic moments, a necessary ingredient for the Kondo effect\cite{perera_spatially_2010,girovsky_long-range_2017,tuerhong_two-dimensional_2018}.

We made a few simplifications in constructing our Hubbard model.
Next-nearest-neighbour hopping was ignored in our Hubbard model because for DCA-Cu this is negligible (Supplementary Fig. \ref*{fig:dft-tb-comparison}).
For simplicity we regarded only on-site Coulomb interactions in the Hubbard model, and although inter-site Coulomb interactions can slightly alter the phase space\cite{fuchs_kagome_2020,wen_interaction-driven_2010,ferhat_phase_2014}, it is generally possible to replace inter-site interactions with an effective on-site interaction\cite{schuler_optimal_2013}.
Given that spin-orbit coupling (SOC) for DCA-Cu is very small\cite{zhang_intrinsic_2016}, we did not consider SOC in our calculations.
Note that when SOC becomes significant (e.g., in systems with heavier elements), it can give rise to effects such as noncollinear magnetism and topological insulator phases\cite{kim_emergent_2020}.
Our DFT calculations also showed some minor changes to the substrate due to the presence of the MOF beyond what we include in the Hubbard model (e.g. graphite in Supplementary Fig. \ref*{fig:tb-fits}), which may lead to differences in the chosen magnetic ordering due to substrate-mediated exchange coupling.
These limitations notwithstanding, our mean-field Hubbard model is adequate for interpreting our DFT results and identifying general trends in phase space.

%Conclusion (Nat Comm has conclusion embedded in the discussion section)
In conclusion, we have shown how choice of substrate controls interaction-induced magnetism in metal-organic frameworks.
MOF-substrate coupling, electron occupation of the MOF, and MOF bandwidth are all critical factors for interaction-induced magnetism (Fig. \ref{fig:summary}).
These can be tuned by choice of substrate (with density of states, lattice mismatch, and work function) or by external controls such as strain or electric field.
The use of electric fields to control magnetism opens the door for electric-field-controlled 2D solid-state devices for spintronics and quantum information technologies.
While our DFT results model DCA-Cu specifically, our Hubbard results are generalisable to any kagome lattice on a substrate.
MOFs (as well as purely organic materials like covalent organic frameworks) benefit from synthesis protocols via supramolecular chemistry, leveraging the versatile toolbox of organic chemistry for tailoring atomic-scale 2D material structure.
We envision experiments where correlated-electron phases of 2D flat-band organic and metal-organic materials -- including the magnetic phases predicted in this work as well as metal-to-Mott-insulator quantum phase transitions -- can be observed via electron transport measurements, optical and photoelectron spectroscopies (including angle-resolved), and scanning probe microscopy techniques (including spin-resolved).
Insights gained here on the physics of substrate-induced magnetism in MOFs can be expected to aid in the rational design of MOFs with exotic properties in realistic device setups.

\clearpage

\section*{Methods}

\subsection*{Density Functional Theory Calculations}

Our density functional theory (DFT) calculations for the 2D MOF DCA-Cu on various substrates were performed using Vienna Ab-initio Simulation Package (VASP)\cite{kresse_efficiency_1996}.
The Perdew-Burke-Ernzerhof (PBE) functional under the generalised gradient approximation (GGA) was used to describe exchange-correlation effects\cite{kresse_ultrasoft_1999}.
Projector augmented wave (PAW) pseudopotentials were used to describe core electrons\cite{blochl_projector_1994,kresse_ultrasoft_1999}.
A semi-empirical potential developed by Grimme (DFT-D3) was used to describe van der Waals forces\cite{grimme_consistent_2010}.
For an accurate description of electron correlations in $d$ electrons, we used Dudarev's implementation of DFT+\textit{U}\cite{dudarev_electron-energy-loss_1998}, which is rotationally invariant and requires only a single free parameter $U$. We used $U=\SI{3}{\eV}$ on DCA-Cu Cu atoms (which is a modest value\cite{wang_dimethylammonium_2013,cockayne_density_2015}) unless otherwise mentioned.
See Supplementary Notes for further details.

For all substrates considered, we used experimental values of the substrate lattice constants\cite{rumble_crc_2019}, as given in Supplementary Table \ref*{tab:substrates}, with DCA-Cu strained to match.
The charge transfer between the substrate and the MOF was determined using Bader analysis\cite{tang_grid-based_2009} and DDEC charge analysis\cite{manz_methods_2011,manz_introducing_2016,limas_introducing_2016,manz_chargemol_2017}.
Two further methods for calculating charge transfer, namely integrating the projected density of states and fitting the kagome band structure, were also used (see Supplementary Notes); they gave values which fell between the Bader and DDEC results.
The local magnetic moments on the MOF were determined by partitioning the spin density then taking the net spin on each DCA molecule.
Both Bader and DDEC analysis methods gave similar results for the magnetic moments.

Visualisation of atomic positions was performed using VESTA\cite{momma_vesta_2011}.
Atomic structures are shown in Supplementary Figure \ref*{fig:structures}.

\subsection*{Mean-Field Hubbard Model}

Our model Hamiltonian for the MOF-substrate system is
\begin{equation}
    \hat{H} = \hat{H}_0 + \hat{H}_U + \hat{H}_{sub} + \hat{H}_{couple}.
    \label{eqn:hamiltonian}
\end{equation}
The first two terms are for the kagome lattice, describing nearest-neighbour hopping and on-site Coulomb interactions respectively\cite{kumar_manifestation_2021}, defined as
\begin{align}
    \hat{H}_0 &= -t \sum_{\langle ij \rangle, \sigma} \cre c{i,\sigma} \ann c{j,\sigma} + \varepsilon,\\
    \hat{H}_U &= U_H \sum_{i} \numn{i,\uparrow} \numn{i,\downarrow} \nonumber \\
    &\approx U_H \sum_i (\numn{i,\uparrow} \mean{\numn{i,\downarrow}}  + \numn{i,\downarrow} \mean{\numn{i,\uparrow}} - \mean{\numn{i,\uparrow}} \mean{\numn{i,\downarrow}}),
\end{align}
where $t$ is the hopping constant, $\varepsilon$ is the on-site energy, $U_H$ is the Coulomb interaction, $\cre{c}{i,\sigma}$ ($\ann{c}{i,\sigma}$) creates (annihilates) an electron with spin $\sigma$ at site $i$, and $\numn{i,\sigma}=\numop{c}{i,\sigma}$ is the density operator.
The Coulomb interactions are solved using the Hartree mean-field approximation.
The substrate is described by the term
\begin{equation}
\hat{H}_{sub} = \sum_{\k,\alpha,\sigma} \epsilon_{\alpha}(\k) \numop{\psi}{\k,\alpha,\sigma},
\end{equation}
where $\cre{\psi}{\k,\alpha,\sigma}$ creates an electron with spin $\sigma$ and momentum $\k$ as a plane wave in substrate band $\alpha$ with dispersion $\epsilon_\alpha(\k)$.
We derive the substrate dispersions from DFT calculations on a substrate slab.
The coupling between the substrate and kagome lattice is given by
\begin{equation}
\hat{H}_{couple} = V \sum_{i,\alpha} (\cre ci \ann{e}{\vr_i,\alpha} + \cre{e}{\vr_i,\alpha} \ann ci),
\end{equation}
where $V$ is the coupling constant and $\cre{e}{\vr,\alpha}$ creates a localised electron in substrate band $\alpha$ at position $\vr$, and $\vr_i$ is the position of kagome lattice site $i$.
We solve this model with a substrate \eqref{eqn:hamiltonian} self-consistently in the grand canonical ensemble.
To get a fixed number of electrons in the kagome lattice, we adjust the relative energy of the kagome lattice and the substrate until the approximately correct electron number is obtained.
Further details of the Hamiltonian are in Supplementary Note \ref*{sec:model}.

\section*{Acknowledgements}

We would like to thank Yuefeng Yin for helpful discussions. B.F. and N.V.M. gratefully acknowledge the computational support from National Computing Infrastructure and Pawsey Supercomputing Facility. B.F. is supported through an Australian Government Research Training Program (RTP) Scholarship. N.V.M. acknowledges funding support from the Australian Research Council (ARC) Centre of Excellence in Future Low-Energy Electronics Technologies (CE170100039). A.S. acknowledges funding support from the ARC Future Fellowship scheme (FT150100426).

\section*{Author Contributions}

B.F. performed the calculations and analyses.
B.F. wrote the custom codes.
A.S. and N.V.M. conceived of and supervised the project.
All authors participated in the writing of this manuscript.

\section*{Competing Interests}

The authors declare no competing interests.

\section*{Data Availability}

The data that support the findings of this study are openly available in figshare at \\ https://doi.org/10.26180/19210632\cite{field_data_2022}

\section*{Code Availability}

The code which implements the mean-field Hubbard model in this work is openly available on GitHub and archived in Zenodo at https://doi.org/10.5281/zenodo.6131285\cite{field_hubbardmf_2022}

\clearpage
% Before submission, we need to split the SI into a separate PDF.

% Tell numbering to enter appendix mode.
\appendix
\renewcommand{\thesubsection}{\arabic{subsection}} % Remove section number from subsection.
\setcounter{figure}{0}
\renewcommand{\thefigure}{\arabic{figure}}
\setcounter{table}{0}
\renewcommand{\thetable}{\arabic{table}}
\setcounter{equation}{0}
\renewcommand{\theequation}{\arabic{equation}}

\maketitle

\section*{Supplementary Information}

\subsection{Coupling-induced antiferromagnetic--ferromagnetic phase transition}\label{sec:afm-fm}

% Paragraph purpose: Discuss how we sometimes get an AFM-FM phase transition.
In our mean-field Hubbard calculations, we observed that for some substrates there was an antiferromagnetic--ferromagnetic transition with increasing $V$, prior to the magnetic--paramagnetic transition.
The kink in the $t=\SI{30}{\milli\eV}$ graphite curve in Fig. \ref*{fig:hubbard-substrate-agvscu}c is an example of this transition.
We also observe a preference for ferromagnetic order when an electric field is applied to DCA-Cu/Cu(111), despite being in a region of phase space where the free-standing system would prefer a frustrated antiferromagnetic configuration.
This ferromagnetism could be due to a substrate-mediated interaction, such at the Ruderman-Kittel-Kasuya-Yosida interaction\cite{van_vleck_note_1962}.
Identifying such interactions is beyond the scope of this study and would require further investigations.

\subsection{Parameters used for density functional theory calculations}

In all our DFT calculations, a \SI{400}{\eV} cut-off was used for the plane wave basis set.
Ionic positions were relaxed until Hellman-Feynman forces were less than \SI{0.01}{\eV\per\angstrom}, using $3\times 3\times 1$ k-point grid for sampling the Brillouin zone, and neglecting spin and $U$.
On substrates we did this relaxation using 1st order Methfessel-Paxton smearing with $\sigma = \SI{0.2}{\eV}$.
For free-standing DCA-Cu we instead used Gaussian smearing with $\sigma = \SI{0.05}{\eV}$ during the relaxation.
Charge and spin density calculations used an $11\times 11\times 1$ k-point grid, energy convergence criterion of \SI{e-6}{\eV}, Blochl tetrahedron interpolation, and dipole corrections.

Bulk substrates were modelled as being three atoms thick.
The bottom layer of atoms was frozen at their bulk positions during structural relaxation.
A layer of passivating hydrogen atoms was applied to the bottom face of metallic substrates (not graphite) to terminate dangling bonds.
The 2D systems were modelled by applying \SI{15}{\angstrom} of vacuum spacing between periodic images.

We used $U=\SI{3}{\eV}$ for all main-text DFT results.
Lack of a quantitative experimental metric sensitive to $U$ such as band gap made fitting to a precise value of $U$ impractical, such that we rely on comparing multiple $U$ values.
We find $U$ provides a small reduction to $t$ and enhancement to the magnetic moments (see Supplementary Fig. \ref{fig:chg-mag-u0} and Supplementary Table \ref{tab:dft-mfh-u-fitting}) but does not otherwise cause any qualitative changes.
This is true even for larger $U$\cite{kumar_manifestation_2021}.
Based on past literature, $U=\SI{3}{\eV}$ is a reasonable lower estimate for Cu atoms in MOFs\cite{wang_dimethylammonium_2013,cockayne_density_2015}, especially considering that DCA-Cu is known to host strong electronic correlations\cite{fuchs_kagome_2020,kumar_manifestation_2021}.

\subsection{Determining charge from the projected density of states}
\label{sec:pdos}

We used Bader and DDEC analysis to obtain estimates for the charge transfer between the MOF and substrate.
To provide a third independent method for determining the charge transfer, we also used the projected density of states (PDOS).
Taking the PDOS projected onto DCA-Cu atomic $spd$ orbitals, we integrated the occupied PDOS and the total PDOS in an energy window around the Fermi energy, then took their ratio.
Multiplying the ratio by two to account for spin degeneracy yields an estimate for the electron occupation of DCA-Cu.
A non-zero background PDOS due to hybridisation with the substrate renders this method sensitive to the window, providing some uncertainty in the precise value of the charge transfer.
We integrated the PDOS from $-1$ to \SI{1}{\eV} to get a central estimate for the charge transfer.
We obtained upper and lower estimates by integrating from $-1$ to \SI{0.5}{\eV} and $-0.5$ to \SI{1}{\eV}.
The energy window was further constrained when shoulders of adjacent bands encroached on these windows.
The charge transfer from the PDOS was found to lie between the values obtained by Bader and DDEC and is shown in Supplementary Fig. \ref{fig:chg-mag-u0}.

\subsection{Determining charge from band structure}\label{sec:charge-from-bands}

For substrates with negligible hybridisation with the MOF, namely DCA-Cu/hBN/Cu(111), we also extracted the charge transfer by fitting a non-spin-polarised band structure.
Here we derive how the position of the Fermi level relative to the band edges gives the electron occupation in an ideal kagome band structure.

For a nearest-neighbour tight-binding model with hopping constant $t$ and neglecting any offset, the kagome band structure is given by
\begin{equation}
\epsilon_{1/2}(\k) = t \left(-1 \pm \sqrt{4 \sum_i \cos^2(\k\cdot\vect{b}_i/2) - 3}\right), \quad
\epsilon_3(\k) = 2t,
\label{eqn:kagome-bands}
\end{equation}
where the normalised lattice vectors can be given by
\begin{equation}
\vect{b}_1 = (\frac{1}{2}, \frac{\sqrt{3}}{2}), \quad
\vect{b}_2 = (1,0), \quad
\vect{b}_3 = \vect{b}_2 - \vect{b}_1.
\end{equation}
Contours of constant energy in k-space for the dispersive bands are given by
\begin{equation}
k_y(k_x,\epsilon) = \frac{2}{\sqrt{3}} \arccos \left( \frac{\epsilon^2/4t^2 + \epsilon/2t - \cos^2(k_x/2)}{\cos(k_x/2)} \right)
\label{eqn:ky}
\end{equation}
The electron occupation is simply the normalised area bounded by a contour of constant energy in the Brillouin zone, times two to account for the two spin channels. Assuming $t>0$,
\begin{equation}
N(\epsilon) = \begin{cases}
0 & \text{ for }\epsilon \le -4t \\
2\times \frac{\sqrt{3}}{2} \frac{1}{(2\pi)^2} \times 4 \int_0^{x_1(\epsilon)} k_y(k_x,\epsilon) dk_x & \text{ for }-4t \le \epsilon \le -2t \\
2 - 2\times \frac{\sqrt{3}}{2} \frac{1}{(2\pi)^2} \times 4 \int_{x_1(\epsilon)}^{x_2(\epsilon)} k_y(k_x,\epsilon) dk_x & \text{ for }-2t \le \epsilon \le -t \\
2 + 2\times \frac{\sqrt{3}}{2} \frac{1}{(2\pi)^2} \times 4 \int_{x_1(\epsilon)}^{x_2(\epsilon)} k_y(k_x,\epsilon) dk_x & \text{ for }-t \le \epsilon \le 0 \\
4 - 2\times \frac{\sqrt{3}}{2} \frac{1}{(2\pi)^2} \times 4 \int_0^{x_1(\epsilon)} k_y(k_x,\epsilon) dk_x & \text{ for }0 \le \epsilon < 2t \\
6 & \text{ for } \epsilon \ge 2t
\end{cases},
\label{eqn:filling}
\end{equation}
where the integration bounds are
\begin{align}
x_1(\epsilon) = 2 \arccos\left(-\frac 12 + \sqrt{\frac{\epsilon^2}{4t^2} + \frac{\epsilon}{2t} +\frac 14}\right), \quad
x_2(\epsilon) = 2 \arccos\left(-\frac 12 - \sqrt{\frac{\epsilon^2}{4t^2} + \frac{\epsilon}{2t} +\frac 14}\right).
\end{align}
The number of electrons is then simply given by \eqref{eqn:filling} at the Fermi level relative to any overall shifting of the bands \eqref{eqn:kagome-bands}. For a given band minimum $\epsilon_{min}$, band maximum $\epsilon_{max}$, and Fermi energy $\epsilon_F$, the units are normalised and given to \eqref{eqn:filling} as
\begin{equation}
N\left( 6 \frac{\epsilon_F - \epsilon_{max}}{\epsilon_{max}-\epsilon_{min}} + 2 \right) \bigg\vert_{t=1}.
\end{equation}
Number of electrons $N$ can be converted to the electron occupation $n$ by dividing by 3, the number of sites per unit cell.

\subsection{Mean-field Hubbard Model}\label{sec:model}

A non-interacting tight-binding model of a lattice with nearest-neighbour hopping has a Hamiltonian of the form
\begin{equation}
\hat{H}_0 = -t \sum_{\langle ij \rangle, \sigma} \cre c{i,\sigma} \ann c{j,\sigma} + \varepsilon,
\end{equation}
where $t$ is the nearest-neighbour hopping constant, $\varepsilon$ is an energy shift, and $\cre c{i,\sigma}$ ($\ann c{i,\sigma}$) creates (annihilates) an electron in the lattice at site $i$ with spin $\sigma$.
Local electron-electron interactions can be included by adding the Hubbard term
\begin{equation}
\hat{H}_U = U_H \sum_{i} \numn{i,\uparrow} \numn{i,\downarrow}
\end{equation}
to the Hamiltonian, where $\numn{i,\sigma}=\numop{c}{i,\sigma}$ is the density operator and $U_H$ is the on-site Coulomb repulsion.
Solving the Hubbard model is difficult.
We apply the Hartee approximation, assuming collinear spins, giving us the mean-field Hamiltonian
\begin{equation}
\hat{H}_U \approx U_H \sum_i (\numn{i,\uparrow} \mean{\numn{i,\downarrow}}  + \numn{i,\downarrow} \mean{\numn{i,\uparrow}} - \mean{\numn{i,\uparrow}} \mean{\numn{i,\downarrow}}).
\end{equation}
See Refs. \cite{kumar_manifestation_2021,field_hubbardmf_2022} for our implementation of this mean-field Hubbard model.

To model the substrate, we assume we have a substrate with some bands with a known dispersion,
\begin{equation}
\hat{H}_{sub} = \sum_{\k,\alpha,\sigma} \epsilon_{\alpha}(\k) \numop{\psi}{\k,\alpha,\sigma},
\end{equation}
where $\cre{\psi}{\k,\alpha,\sigma}$ creates an eigenstate of the substrate with wavevector $\k$, band index $\alpha$, spin $\sigma$, and eigenenergy $\epsilon_{\alpha}(\k)$. The eigenenergies may be derived from a simple tight-binding model, or they may be directly calculated by DFT for a given substrate.
In our calculations, we used substrate dispersions obtained from DFT calculations using a single substrate unit cell and $48\times 48\times 1$ k-points, without spin or $U$, but otherwise with the same parameters as the other DFT calculations, including the slab thickness.
We assume the substrate eigenstates are simple plane waves (that is, the substrate is homogeneous in real-space).
This gives the relation between momentum and position eigenstates
\begin{equation}
\bra{0} \ann{e}{\vr,\alpha} \cre{\psi}{\k,\beta} \ket{0} = \frac{1}{\sqrt{N N_R}} e^{i \k\cdot\vr} \delta_{\alpha,\beta},
\end{equation}
where $N$ is the number of substrate cells which fit within the lattice cell, $N_R$ is the number of lattice cells in the entire system, and $\cre{e}{\vr,\alpha}$ creates an electron in the substrate in band $\alpha$ at the point $\vr$.
As a consequence of the substrate eigenstates being homogeneous and to simplify calculations we also assume that there is no Coulomb interaction within the substrate.
We also assume that the substrate is commensurate with the lattice supercell and that $N$ primitive unit cells of the substrate fit within the supercell of the lattice.
This means for each band $\alpha$ in a single substrate cell we have $N$ copies of that band in the supercell Brillouin zone, each shifted by a reciprocal lattice vector.

Finally, we assume coupling between the substrate and lattice is point-like, which gives us the coupling term
\begin{equation}
\hat{H}_{couple} = \sum_{i,\alpha} V_\alpha (\cre ci \ann{e}{\vr_i,\alpha} + \cre{e}{\vr_i,\alpha} \ann ci),
\end{equation}
where $V_\alpha$ is the coupling constant between the lattice and the substrate band $\alpha$, and $\vr_i$ is the position of lattice site $i$.
To perform calculations in a periodic system, we convert the lattice basis into a Bloch basis,
\begin{equation}
\cre{\phi}{\k,i,\sigma} = \frac{1}{\sqrt{N_R}} \sum_{\vect{R}} e^{i\k\cdot\vect{R}} \cre{c}{\vect{R},i,\sigma},
\end{equation}
where $\vect{R}$ is a lattice vector. The conversion of $\hat{H}_0$ and $\hat{H}_U$ occur by the normal means. The coupling term has matrix elements given by
\begin{equation}
\bra{0} \ann{\phi}{\k,i,\sigma} \hat{H}_{couple} \cre{\psi}{\k+\vect{G},\alpha,\sigma} \ket{0} = \frac{V_\alpha}{\sqrt{N}} e^{i(\k + \vect{G})\cdot \vr_i},
\end{equation}
where $\vect{G}$ is a reciprocal lattice vector and $\vr_i$ here refers to the position within the home unit cell.

The overall Hamiltonian for the Hubbard model with the substrate is
\begin{equation}
\hat{H} = \hat{H}_0 + \hat{H}_{sub} + \hat{H}_{couple} + \hat{H}_U.
\end{equation}
For concreteness, consider a single kagome unit cell and a single substrate band. In matrix form the Hamiltonian for a single spin channel is
\begin{equation}
{H}_{\sigma}(\k) = \begin{pmatrix}
\varepsilon + U_H n_{1,-\sigma} & -t(1+e^{-i\k\cdot\vect{b}_1}) & -t(1+e^{-i\k\cdot\vect{b}_2}) & \tfrac{V}{\sqrt{N}} & \tfrac{V}{\sqrt{N}} & \dots \\
-t(1+e^{i\k\cdot\vect{b}_1}) & \varepsilon + U_H n_{2,-\sigma} & -t(1+e^{-i\k\cdot\vect{b}_3}) & \tfrac{V}{\sqrt{N}} e^{i(\k+\vect{G}_1)\cdot\vect{b}_1/2} & \tfrac{V}{\sqrt{N}} e^{i(\k+\vect{G}_2)\cdot\vect{b}_1/2} & \dots \\
-t(1+e^{i\k\cdot\vect{b}_2}) & -t(1+e^{i\k\cdot\vect{b}_3}) & \varepsilon + U_H n_{3,-\sigma} & \tfrac{V}{\sqrt{N}} e^{i(\k+\vect{G}_1)\cdot\vect{b}_2/2} & \tfrac{V}{\sqrt{N}} e^{i(\k+\vect{G}_2)\cdot\vect{b}_2/2} & \dots \\

\tfrac{V}{\sqrt{N}} & \tfrac{V}{\sqrt{N}} e^{-i(\k+\vect{G}_1)\cdot\vect{b}_1/2} & \tfrac{V}{\sqrt{N}} e^{-i(\k+\vect{G}_1)\cdot\vect{b}_2/2} & \epsilon(\k+\vect{G}_1) & 0 & \\
\tfrac{V}{\sqrt{N}} & \tfrac{V}{\sqrt{N}} e^{-i(\k+\vect{G}_2)\cdot\vect{b}_1/2} & \tfrac{V}{\sqrt{N}} e^{-i(\k+\vect{G}_2)\cdot\vect{b}_2/2} & 0 & \epsilon(\k+\vect{G}_2) &  \\
\vdots & \vdots & \vdots & & & \ddots \\
\end{pmatrix},
\end{equation}
where $n_{i,\sigma} \equiv \mean{\numn{i,\sigma}}$, which are found self-consistently from the occupied eigenstates of ${H}_\sigma$.
The total energy is the sum of occupied eigenenergies from each spin channel plus $-U\sum_i n_{i,\uparrow} n_{i,\downarrow}$.
Extra substrate bands can be added by concatenating the blocks for the coupling term and substrate dispersion to $H_\sigma$.

We solve this substrate model in the grand canonical ensemble, meaning we fix the chemical potential $\mu$.
This affects the occupation of single-particle eigenstates and thus the electron density in the self-consistency loop. We also subtract $\mu N_e$ from the total energy (where $N_e$ is the total electron number).
Using the grand canonical ensemble simulates the large reservoir of electrons which a real substrate would provide without having to simulate a large number of bulk bands.
Where a fixed number of electrons in the lattice was desired, we incrementally changed the relative alignment of the lattice and substrate bands until the electron occupation was close to the desired value.
The chemical potential of the substrate was kept constant.
Conversely, free-standing calculations were performed in the canonical ensemble where the number of electrons is kept constant.
To get the spin configurations, we initially compare multiple different initial spin configurations and keep the lowest energy configuration.
For calculations with increasing substrate coupling we use the spin configuration from the previous value of coupling as the trial spin configuration to speed up convergence.
At each point we also compared to a fully paramagnetic configuration.
All calculations with a substrate considered a single kagome unit cell and a number of substrate unit cells comparable to the DFT calculations.
In single kagome unit cell calculations, we considered a $25\times 25$ k-point grid.
Kagome supercell calculations (used for free-standing systems) had the k-point grid scaled to keep a similar k-point density.

\clearpage

\begin{table}
    \centering
    \begin{tabular}{c|ccc|ccc}
         Charge & \multicolumn{3}{c|}{$U=\SI{0}{\eV}$} & \multicolumn{3}{c}{$U=\SI{3}{\eV}$}  \\
         ($e$/unit cell) & $t$ (meV) & $U_H/t$ & $U_H$ & $t$ (meV) & $U_H/t$ & $U_H$ \\
         \hline
         -1 & 53.6 & 5.31 & 285 & 49.2 & 5.90 & 290 \\
         -0.75 & 53.6 & 5.43 & 291 & 49.2 & 5.93 & 292 \\
         -0.5 & 53.6 & 5.63 & 302 & 49.2 & 5.86 & 288 \\
         -0.25 & 53.6 & 5.63 & 302 & 49.2 & 5.89 & 290 \\
         0 & 53.6 & $<6$ & $<322$ & 49.2 & 6.10 & 300 \\
         +0.25 & 53.6 & $\sim 6$ & $\sim 322$ & 49.2 & 6.50 & 320 \\
         +0.5 & 53.6 & 6.42 & 344 & 49.2 & 7.05 & 347 \\
         +0.75 & 53.6 & 6.25 & 335 & 49.2 & 6.79 & 334 \\
    \end{tabular}
    \caption{Fitting of Hubbard model parameters to DFT+$U$ calculations of free-standing DCA-Cu.
    Hopping $t$ is determined from the bandwidth of a non-spin-polarised band structure.
    $U_H/t$ is determined by fitting the magnetization density of the mean-field Hubbard model to the DFT results on each DCA molecule.
    $U=0$ results at charges of 0 and +0.25 are difficult to fit to the Hubbard model because they had vanishing magnetization densities.
    Comparing DFT+$U$ results for $U=0$ and $U=\SI{3}{\eV}$, we find that $U_H$ is mostly unchanged between the two, with $t$ changing instead.
    This is an intuitive result considering the Wannier orbitals\cite{fuchs_kagome_2020}.
    The Cu orbitals have only a small contribution to the Wannier orbitals, but they dominate the overlap between adjacent Wannier orbitals, so $U$ on the Cu atoms will affect $t$ more than $U_H$.}
    \label{tab:dft-mfh-u-fitting}
\end{table}

\begin{table}
    \centering
    \begin{tabular}{c|cccc|ccc}
        System & Config. & Energy (meV) & $\sqrt{\langle m^2 \rangle}$ & $\langle m \rangle$ & DCA 1 ($\mu_{\rm B}$) & DCA 2 ($\mu_{\rm B}$) & DCA 3 ($\mu_{\rm B}$) \\
        \hline
        Free (Neutral) & AFM & 0.01 & 0.247 & 0 & 0.175 & 0.175 & -0.349 \\
                       & AFM & 0 & 0.249 & 0 & 0.305 & -0.305 & 0 \\
                       & PM & 0.95 & 0 & 0 & 0 & 0 & 0 \\
        \hline
        Ag(111) & AFM & 0 & 0.241 & -0.052 & -0.218 & -0.218 & 0.283 \\
                & AFM & 0.94 & 0.194 & 0 & -0.237 & 0.237 & 0 \\
                & PM & 1.46 & 0 & 0 & 0 & 0 & 0 \\
        \hline
        hBN/Cu(111) & AFM & 0 & 0.572 & 0.178 & 0.521 & 0.521 & -0.663 \\
                    & FM & 23.0 & 0.058 & 0.058 & 0.058 & 0.058 & 0.058 \\
                    & PM & 23.9 & 0 & 0.001 & 0 & 0 & 0 \\
        \hline
        Graphite & PM & 0 & 0.005 & 0.002 & 0.003 & 0.006 & -0.005 \\
                 & AFM & 0.2 & 0.026 & 0.012 & -0.019 & 0.037 & 0.019 \\
        \hline
        Cu(111) & FM & 0 & 0.290 & 0.280 & 0.290 & 0.290 & 0.290 \\
        E=\SI{0.5}{\eV\per\angstrom}
                & AFM & 0.4 & 0.231 & 0.099 & 0.254 & 0.254 & -0.179 \\
                & AFM & 1.2 & 0.204 & 0 & 0.249 & -0.249 & 0 \\
                & PM & 1.4 & 0.001 & 0.001 & 0.001 & 0.001 & 0.001 \\
        \hline
        hBN/Cu(111)
                    & FM & 0 & 0.149 & 0.149 & 0.149 & 0.149 & 0.149 \\
        E=\SI{-0.5}{\eV\per\angstrom}
                    & PM & 0.4 & 0 & 0 & 0 & 0 & 0 \\
        \hline
        Graphite
                & AFM & 0 & 0.115 & -0.008 & -0.171 & 0.073 & 0.074 \\
        2.7\% Strain
                & PM & 0.3 & 0 & 0 & 0.001 & 0 & 0
    \end{tabular}
    \caption{Relative energies of different spin configurations of DCA-Cu, calculated by DFT.
    Energies are per unit cell and relative to the lowest energy configuration.
    Local magnetic moments $\sqrt{\langle m^2 \rangle}$ and net magnetic moments $\langle m \rangle$ are measured per DCA.
    Any difference between $\langle m \rangle$ and the average of the magnetic moments on DCA is due to magnetic moments on Cu or the substrate.}
    \label{tab:spin-summary}
\end{table}

\begin{table}
	\centering
	\begin{tabular}{ccccc}
		Substrate & Lattice (\si{\angstrom}) & Strain & Work function (\si{\eV}) & DOS (atom$^{-1}$\si{\per\eV}) \\
		\hline
		No substrate & 20.022 & 0.0\% & 3.95 (This work) & -- \\
		Ag(111) ($7\times 7$) & 20.223 & 1.0\% & 4.74\cite{rumble_crc_2019} & 0.27 \\
		Cu(111) ($8\times 8$) & 20.432 & 2.0\% & 4.94\cite{rumble_crc_2019} & 0.30 \\
		hBN/Cu(111) ($8\times 8$) & 20.432 & 2.0\% & 3.62\cite{joshi_boron_2012} & 0 \\
		Al(111) ($7\times 7$) & 20.045 & 0.1\% & 4.26\cite{rumble_crc_2019} & 0.41 \\
		Au(111) ($7\times 7$) & 20.186 & 0.8\% & 5.31\cite{rumble_crc_2019} & 0.28 \\
		Graphite ($8\times 8$) & 19.690 & $-1.7$\% & 4.7\cite{rutkov_graphene_2020} & 0.01* \\
	\end{tabular}
	\caption{Physical properties of the substrates.
	Lattice constants are those used in calculations.
	Strain is the strain of DCA-Cu relative to the free-standing system.
	Work functions and bulk densities of states (DOS) are provided to illustrate the range of different substrate parameters considered and as indicative of possible charge transfer and hybridisation, respectively.
	DOS is taken at the Fermi level and calculated using DFT with a $48\times 48\times 48$ k-point grid and no spin or $U$, except graphite which used a $48\times 48\times 18$ k-point grid due to different unit cell vectors and had the DOS averaged over \SI{+-0.2}{\eV} due to the Dirac point.
	Note that this bulk DOS is slightly different to the DOS of a thin slab used in Fig. \ref*{fig:hubbard-substrate-agvscu}c.}
	\label{tab:substrates}
\end{table}

\begin{figure}
	\centering
	\includegraphics{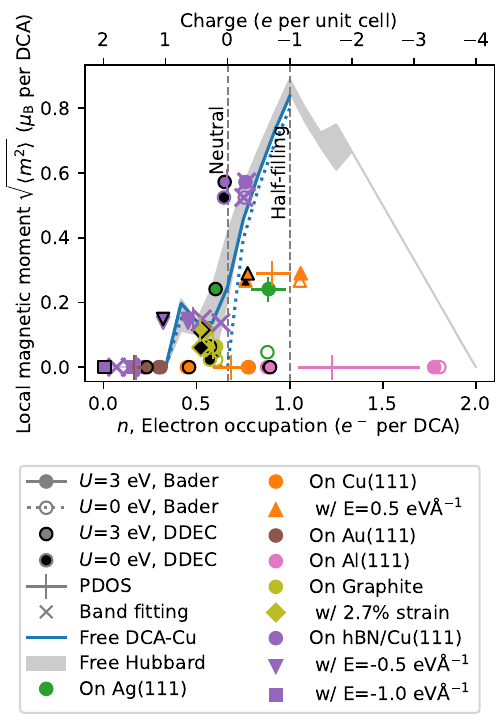}
	\caption{Summary of charge transfer and local magnetic moments of DCA-Cu (free-standing and on substrates) under different conditions, calculated using DFT.
	Data is as in Fig. \ref*{fig:chg-mag-summary}, but $U=0$ data is included here, as well as integrating the projected density of states (PDOS) (Supplementary Note \ref{sec:pdos}) and fitting to non-spin-polarised or ferromagnetic kagome band structures (Supplementary Note \ref{sec:charge-from-bands}) to obtain other measures of the charge transfer.
	The inclusion of $U$ in DFT+$U$ causes a slight enhancement in the magnetic moments but otherwise does not cause any qualitative changes.
	The exception is Ag(111), where $U$ brings the system from nearly non-magnetic to noticeably magnetic.
	This is because DCA-Cu/Ag(111) sits near the phase boundary between non-magnetic and magnetic so is sensitive to small changes.}
	\label{fig:chg-mag-u0}
\end{figure}

\begin{figure}
	\includegraphics{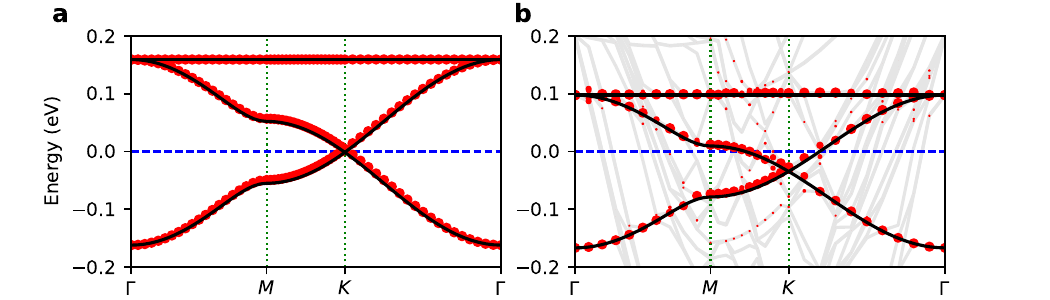}
	\caption{Comparison between non-spin-polarised DFT ($U=0$) band structures of (\textbf{a}) free-standing DCA-Cu and (\textbf{b}) DCA-Cu/hBN/Cu(111) and a kagome tight-binding model.
	Red circles mark DFT projection onto DCA-Cu.
	Black lines are a nearest-neighbour tight-binding model fitted to the bandwidth at the Gamma point.
	Grey lines are substrate bands from DFT.
	Free-standing DCA-Cu is well described by a nearest-neighbour tight-binding model (with $t=\SI{53.6}{\milli\eV}$).
	DCA-Cu/hBN/Cu(111) is well-approximated as free-standing (with $t=\SI{44.1}{\milli\eV}$).
	Note that results with $U$ have slightly lower values of $t$.}
	\label{fig:dft-tb-comparison}
\end{figure}

\begin{figure}
	\includegraphics{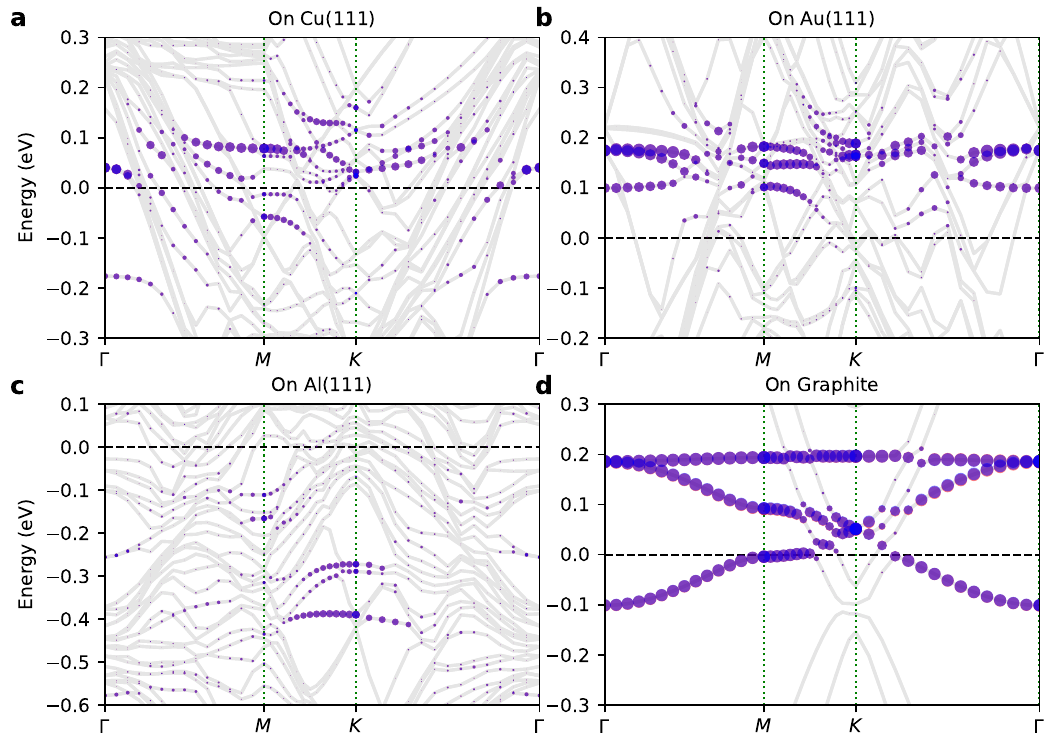}
	\caption{Band structures of (\textbf{a}) DCA-Cu/Cu(111), (\textbf{b}) DCA-Cu/Au(111), (\textbf{c}) DCA-Cu/Al(111), and (\textbf{d}) DCA-Cu/Graphite, calculated by DFT.
	Circle radii are proportional to the projection onto DCA-Cu.
	All the bands here are spin degenerate, indicating an absence of magnetic moments when DCA-Cu is on these substrates.}
	\label{fig:misc-bands}
\end{figure}

\begin{figure}
		\centering
       \includegraphics{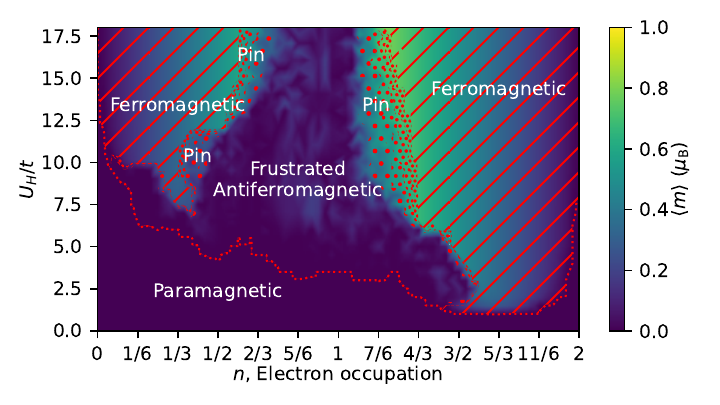}
       \caption{Net magnetic moment of the Hubbard model without a substrate, with respect to interaction parameter $U_H$ and electron occupation $n$, in a 6-by-6 supercell.
       Compare with Figure \ref*{fig:mfh}b for the local magnetic moments $\sqrt{\langle m^2 \rangle}$.
   	   The normalised difference between the net magnetic moment and $\sqrt{\langle m^2 \rangle}$ is used to broadly categorise different phases.}
       \label{fig:hubbard-un-mag}
\end{figure}

\begin{figure}
	\centering
	\includegraphics{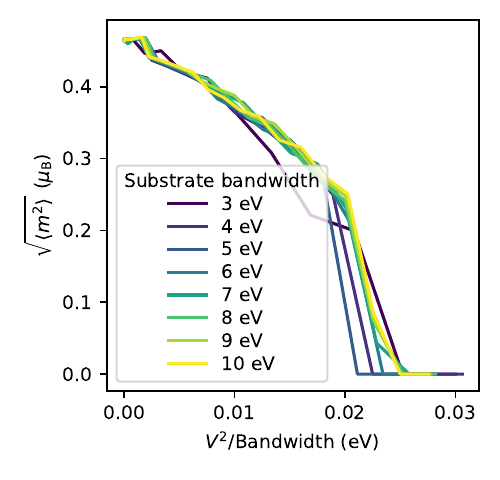}
	\caption{Comparison between Hubbard models with substrates which vary only by their bandwidth and DOS.
	The substrate is a $5\times 5$ supercell of a triangular nearest-neighbour single-orbital tight-binding model.
	The kagome lattice uses $t=\SI{40}{\milli\eV}$, $U_H=\SI{300}{\milli\eV}$ and $n\approx 2/3$.
	The x-axis plots coupling squared divided by substrate bandwidth.
	Because the bandwidth is the only feature which differs between these substrates, substrate DOS is inversely proportional to bandwidth so the x-axis is $V^2\rho$ up to some scale factor.
	Changing the substrate DOS causes only a predictable rescaling of the effective coupling.}
	\label{fig:hubbard-triangle-dos}
\end{figure}

\begin{figure}
		\centering
       \includegraphics{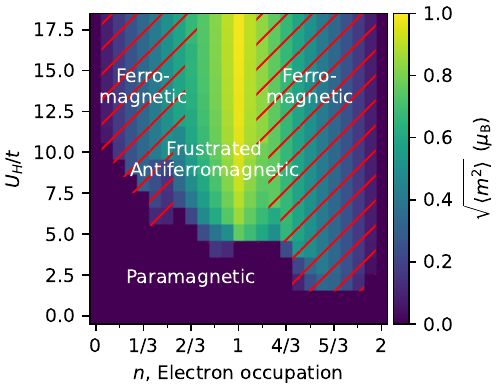}
       \includegraphics{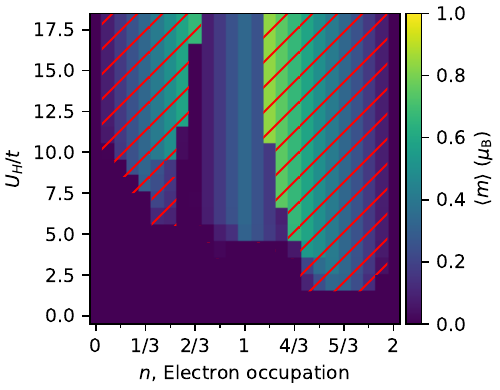}
       \caption{Phase diagram of the Hubbard model without a substrate, with respect to interaction parameter $U_H$ and electron occupation $n$, in a single unit cell.
       Left: Magnitude of local magnetic moments.
       Right: Net magnetic moment.
       Note the similar trends to the supercell phase diagram (Fig. \ref*{fig:hubbard-un}), but with no intermediate `pin' phase because the unit cell is too small to describe that phase.
       We also get a consistent net magnetic moment in much of the frustrated antiferromagnetic phase, making it more closely resemble ferrimagnetism, unlike the supercell (Supplementary Figure \ref{fig:hubbard-un-mag}) where the net magnetic moment is approximately zero (noise notwithstanding).}
       \label{fig:hubbard-un-single}
\end{figure}

\begin{figure}
	\centering
	\includegraphics{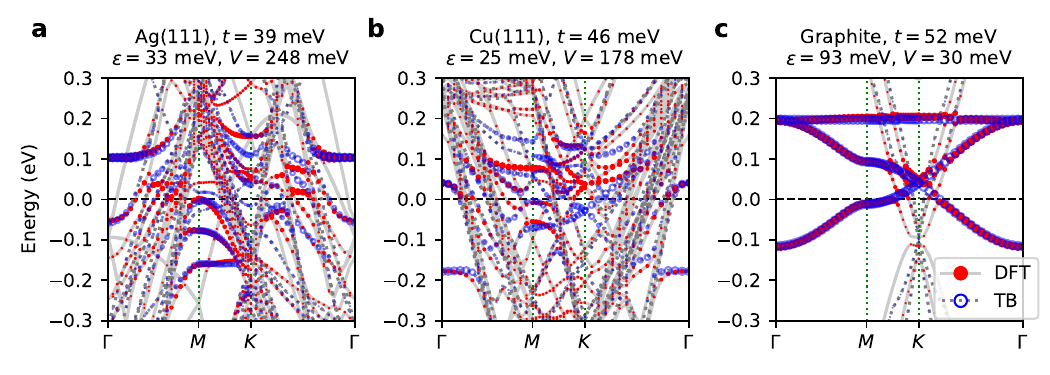}
	\caption{Fitting of tight-binding parameters to DFT results with a substrate.
	DFT results are non-spin-polarised, $U=0$.
	Tight-binding results are with the Hubbard model with $U_H=0$.
	Initial fit parameters were obtained manually, then refined by convolving the projected band structures with a Lorentzian (width \SI{5}{\milli\eV}) then maximising their cross-correlation.
	While quantitative features differ, the model captures qualitative features, especially for the metallic substrates.
	The graphite results (\textbf{c}), however, do not properly reflect the coupling in the DFT results, possibly due to spatially extended coupling (i.e. not point-like) or MOF-induced substrate reconstruction.}
	\label{fig:tb-fits}
\end{figure}

\begin{figure}
	\centering
	\includegraphics{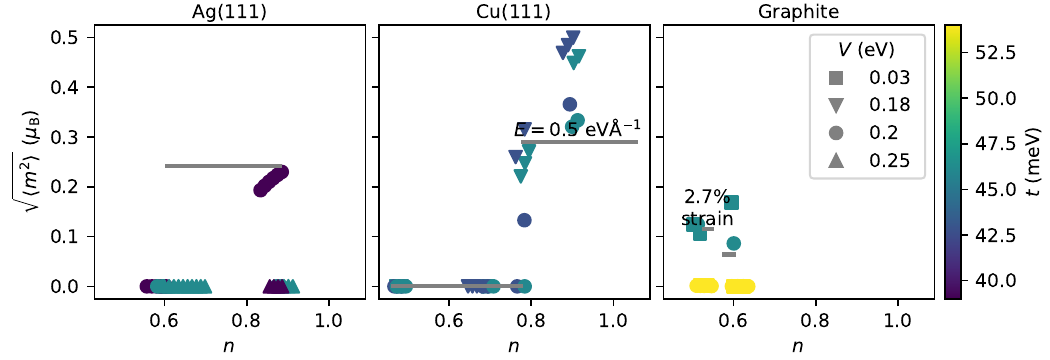}
	\caption{Local magnetic moments in the Hubbard model on a substrate at parameters derived from DFT results.
	$U_H=\SI{0.3}{\eV}$ is used.
	Grey bars are the charge and magnetic moments obtained from DFT calculations on the same substrate, with the annotated strain or electric field, otherwise without applied strain or electric field.
	As illustrated, the Hubbard model reproduces the DFT-derived local magnetic moments in a portion of the Hubbard model parameter space.
    Although not always a perfect match, this affirms that our simple model is appropriate for modelling the fundamental physics of this system.}
	\label{fig:mfh-chosen-parameters}
\end{figure}

\begin{figure}
	\centering
	\includegraphics{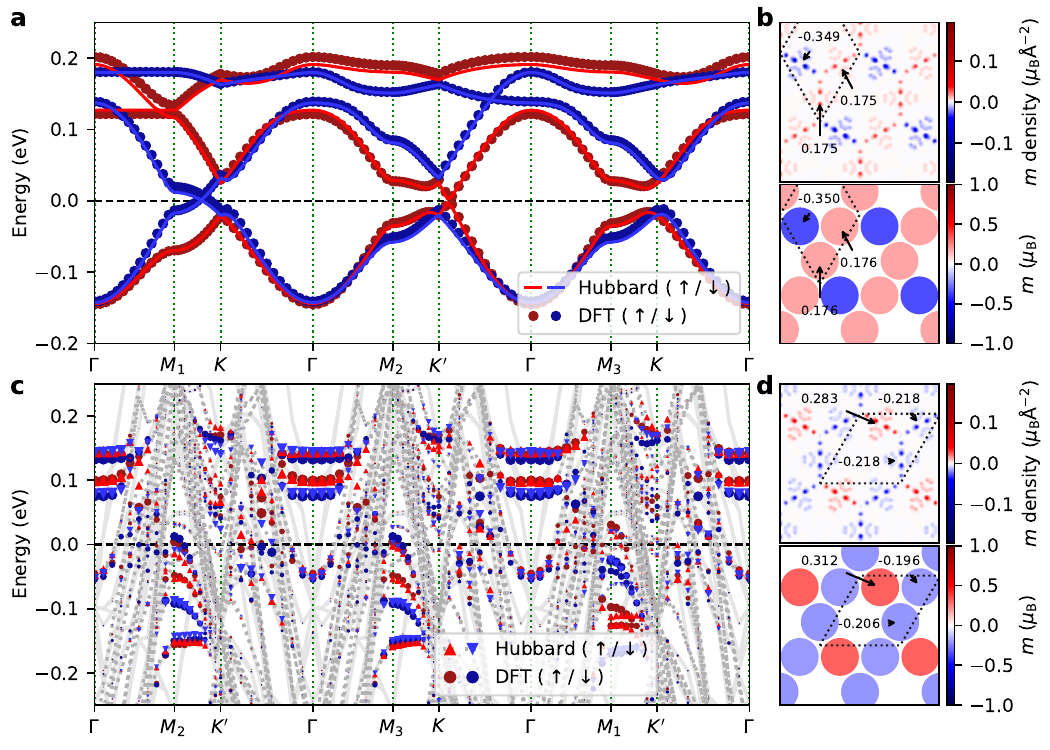}
	\caption{Comparison of band structures and magnetization densities calculated by DFT+$U$ and the mean-field Hubbard model.
		\textbf{a} Band structure of free-standing DCA-Cu (as in Figure \ref*{fig:bands}a).
		Hubbard parameters are $t=\SI{49.2}{\milli\eV}$ and $U_H=\SI{300}{\milli\eV}$ (Supplementary Table \ref{tab:dft-mfh-u-fitting}).
		Hopping $t$ was obtained from the bandwidth of a non-spin-polarised DFT band structure.
		The Coulomb repulsion $U_H$ was obtained by fitting to the magnetization density of free-standing DCA-Cu.
		No fitting to the spin-polarised band structure was performed.
		\textbf{b} Magnetization density of free-standing DCA-Cu from DFT+$U$ (top) (as in Figure \ref*{fig:bands}d) and the mean-field Hubbard model (bottom), with parameters as in \textbf{a}.
		Note that the DFT results are an area-density, while the Hubbard results are for the total magnetization across a whole molecule.
		Annotations give the value of the total magnetization on each molecule to allow for a direct comparison.
		The dashed rhombus is the primitive unit cell.
		\textbf{c} Band structure of DCA-Cu/Ag(111) (as in Figure \ref*{fig:bands}b).
		Hubbard parameters are $t=\SI{37}{\milli\eV}$, $U_H=\SI{360}{\milli\eV}$, $\varepsilon=\SI{-110}{\milli\eV}$, and $V=\SI{248}{\milli\eV}$, and the spin density was relaxed from the DFT+$U$ spin density.
		These parameters gave $\sqrt{\langle m^2 \rangle}=0.244$, $\langle m \rangle=-0.029$, and $n=0.87$.
		$V$ was taken from Supplementary Figure \ref{fig:tb-fits}.
		$t$ was reduced slightly from the $U=0$ results in Supplementary Figure \ref{fig:tb-fits}, following the reduction in $t$ observed in Supplementary Table \ref{tab:dft-mfh-u-fitting}, with some slight adjustment to fit the spin-polarised band structure.
		$U_H$ and $\varepsilon$ were manually chosen to give $\sqrt{\langle m^2 \rangle}$ and $n$ values approximately corresponding with the DFT+$U$ results.
		\textbf{d} Magnetization density of DCA-Cu/Ag(111) (as in Figure \ref*{fig:bands}e), with Hubbard parameters as in \textbf{c}.\\
		We emphasise that, besides a slight (5\%) adjustment to $t$ for DCA-Cu/Ag(111) in line with using a finite $U$, no fitting to the spin-polarised band structures was used.
		Instead, all parameters were obtained by fitting to non-spin-polarised band structures (for $t$ and $V$) and magnetization densities (for $U_H$ and $\varepsilon$).
		These quantities are simpler and easier to compute than the intricate and complicated spin-polarised band structure.
		Overall, these band structures show remarkable agreement between the DFT results and the Hubbard model.
		This indicates that the Hubbard model can reproduce DFT results in these systems, allowing for interpretation of the DFT results.}
	\label{fig:dft-mfh-bands}
\end{figure}

\begin{figure}
	\centering
	\begin{tabular}{cc}
		Ag(111) & Cu(111) \\
		\includegraphics[width=0.42\linewidth]{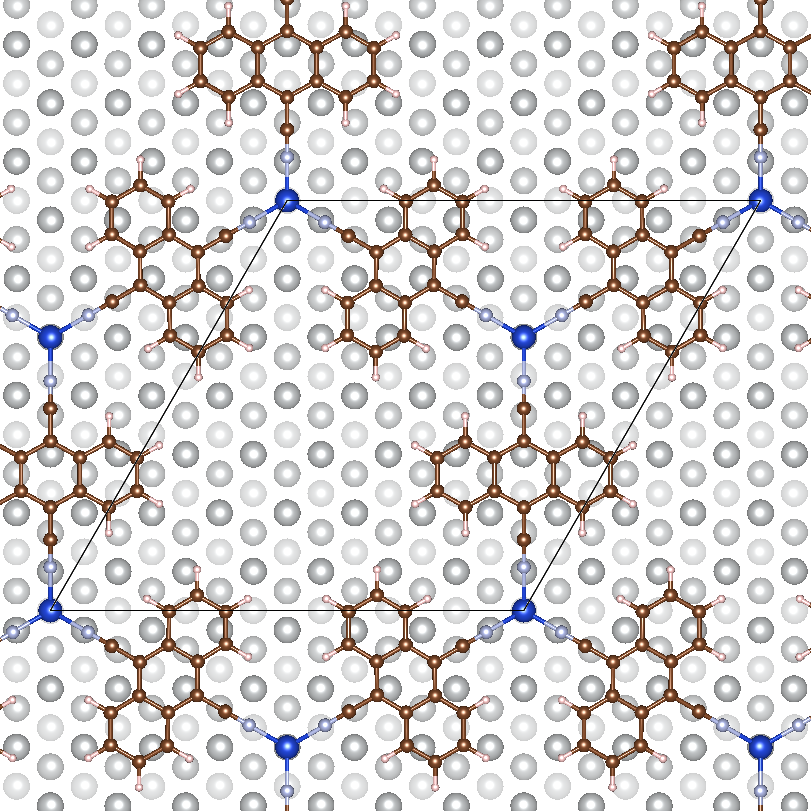} &
		\includegraphics[width=0.42\linewidth]{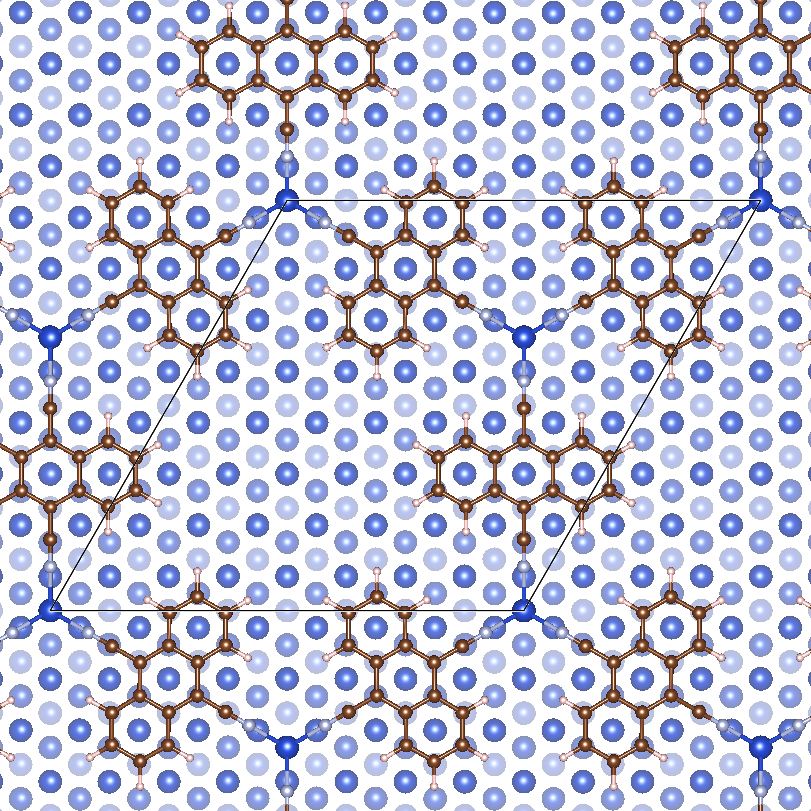} \\
		Al(111) & Au(111) \\
		\includegraphics[width=0.42\linewidth]{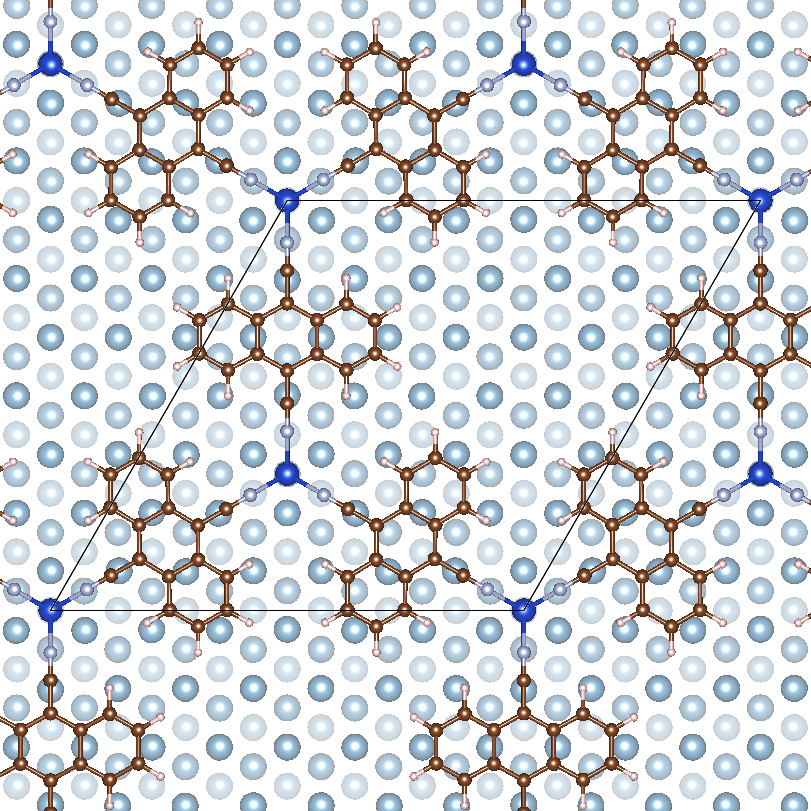} &
		\includegraphics[width=0.42\linewidth]{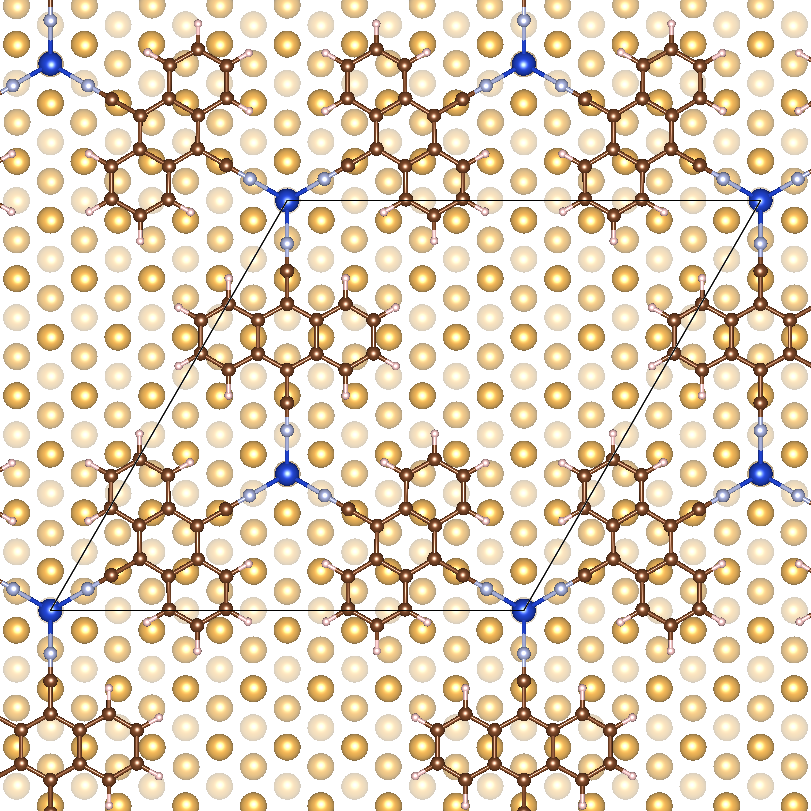} \\
		Graphite & hBN/Cu(111) \\
		\includegraphics[width=0.42\linewidth]{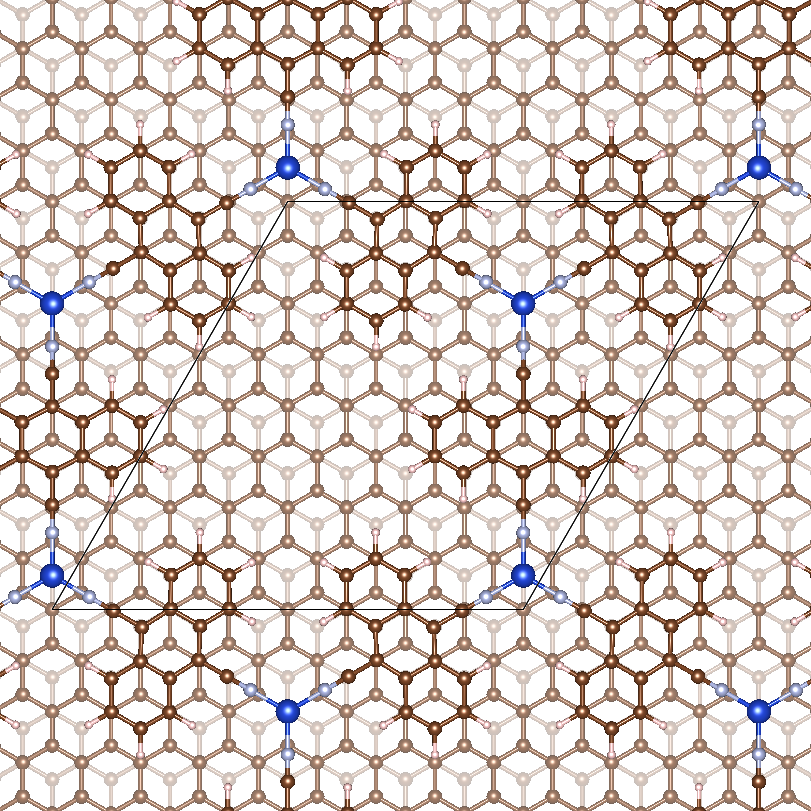} &
		\includegraphics[width=0.42\linewidth]{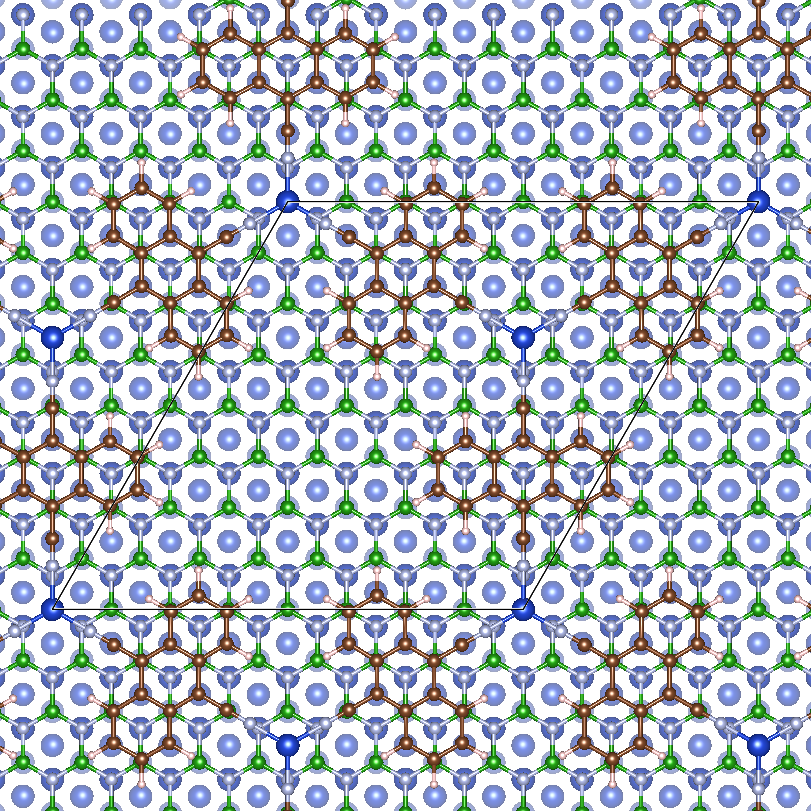}
	\end{tabular}
	\includegraphics[scale=0.25]{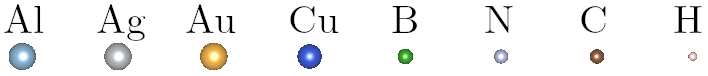}
	\caption{Top-down views of structures used for DCA-Cu on substrates.
	Lighter-coloured atoms are deeper.
	The black rhombus outlines the unit cell.}
	\label{fig:structures}
\end{figure}

\end{document}